\documentclass[aps,prd,preprintnumbers,nofootinbib,showpacs,eqsecnum,12pt]{revtex4-1}
\usepackage{amsmath}         
\usepackage{amssymb}          
\usepackage[pdftex]{graphicx}	
\usepackage[pdftex]{hyperref} 
\usepackage[utf8x]{inputenc}
\usepackage{color}
\usepackage{ulem}
\usepackage{afterpage}

\newcommand{\ord}[1]{\mathcal{O}\left({#1}\right)}

\def\gsim{\raise0.3ex\hbox{$\;>$\kern-0.75em\raise-1.1ex\hbox{$\sim\;$}}}
\def\lsim{\raise0.3ex\hbox{$\;<$\kern-0.75em\raise-1.1ex\hbox{$\sim\;$}}}

\begin{document}

\title{Resolving a challenging supersymmetric low-scale seesaw scenario at the ILC}

\author{J. Masias}
\email[E-mail: ]{j.masias@pucp.edu.pe}
\author{N. Cerna-Velazco}
\email[E-mail: ]{n.cerna@pucp.edu.pe}
\author{J. Jones-P\'erez}
\email[E-mail: ]{jones.j@pucp.edu.pe}
\affiliation{
Secci\'on F\'isica, Departamento de Ciencias, Pontificia Universidad Cat\'olica del Per\'u, Apartado 1761, Lima, Peru
}
\author{W. Porod}
\email[E-mail: ]{porod@physik.uni-wuerzburg.de}
\affiliation{
Institut f\"ur Theoretische Physik und Astrophysik, Uni W\"urzburg
}

\begin{abstract}
We investigate a scenario inspired by natural supersymmetry, where neutrino data is explained within a low-scale seesaw scenario. For this the Minimal Supersymmetric Standard Model is extended by adding light right-handed neutrinos and their superpartners, the R-sneutrinos. Moreover, we consider the lightest neutralinos to be Higgsino-like. We first update a previous analysis and assess to which extent does existing LHC data constrain the allowed slepton masses. Here we find scenarios where sleptons with masses as low as 175 GeV are consistent with existing data. However, we also show that the upcoming run will either discover or rule out sleptons with masses of 300 GeV, even for these challenging scenarios.

We then take a scenario which is on the borderline of observability of the upcoming
LHC run assuming a luminosity of 300 fb$^{-1}$. We demonstrate that a prospective
international $e^+ e^-$ linear collider with a center of mass energy of 1 TeV will
be able to discover sleptons in scenarios which are difficult for the LHC. Moreover,
we also show that a measurement of the spectrum will be possible within 1-3 percent
accuracy.
\end{abstract}

\maketitle

\section{Introduction}

Particle physics faces currently a somewhat paradoxical situation: on the one hand we have 
the Standard Model (SM) of particle physics predicting a wealth of phenomena which
have been scrutinized and confirmed by various experiments. An important player
is here the Large Hadron Collider (LHC) which delivered a huge amount of data
in the last year and is currently preparing for a new run. On the other hand,
there is direct evidence that the SM needs to be extended, the most prominent ones being neutrino masses and mixing~\cite{GonzalezGarcia:2007ib,Schwetz:2008er,Gariazzo:2018pei}, as well as dark matter~\cite{Bertone:2004pz,Profumo:2019ujg}. Moreover, the structure of
the SM also suggests that it should be considered as an effective theory to be embedded in
a more fundamental one at high energy scales. For example, the product structure of the gauge group $SU(3)_C \times SU(2)_L \times U(1)_Y$ hints toward an embedding in a larger group like $SU(5)$ or $SO(10)$. However, if one evolves the gauge coupling via renormalization group equations (RGEs) up to higher scales, they do not unify if one insists of using the SM particle content only~\cite{Amaldi:1991cn,Langacker:1991an,Ellis:1990wk}. The Higgs mass term of the SM is the only relevant operator in the sense that it is sensitive to physics at arbitrary large energy scales. Here the question arises how to stabilize the Higgs mass at the electroweak scale.

Supersymmetry (SUSY) is very likely up to now the most studied extension of the SM, one of the reasons being that it addresses the last two issues. Moreover, it yields also a possible candidate for the observed dark matter \cite{Ellis:1983ew}. In view of neutrino physics, the minimal model needs to be extended, which can be achieved for example by a supersymmetric variant of the seesaw mechanism.

It was anticipated that SUSY should be discovered relatively fast at the LHC. 
However, after several years of running, no sign of physics beyond the SM has been
observed, with the potential exception of some anomalies related the lepton universality in the B-meson sector~\cite{deSimone:2020kwi}. This clearly shows that original vanilla forms of SUSY, like the constrained minimal supersymmetric standard model (CMSSM) or gravity mediated SUSY breaking (GMSB), are not realized in nature. However, this by far does not exclude SUSY per se, and it was early on noticed that there are several scenarios which can potentially evade detection at the LHC for a long time~\cite{LeCompte:2011fh,Fan:2012jf,Kribs:2012gx,Cahill-Rowley:2014twa,Carpenter:2020fnh}.

In a previous work we investigated a supersymmetric model where neutrino masses and mixing are generated via the inclusion of a low scale seesaw mechanism~\cite{Cerna-Velazco:2017cmn}. We demonstrated that in certain parts of the parameter space sleptons with masses as low as $\sim150$~GeV
were still consistent with LHC data. This is considerably lower than current slepton mass bounds within the MSSM \cite{Aad:2014vma,Sirunyan:2018nwe,Aad:2019qnd,Aad:2019vnb,Aad:2019byo}. In this
region the SUSY partners of the right-handed neutrinos, the R-sneutrinos, are
the lightest supersymmetric particles (LSPs). In addition, the sleptons are lighter than both neutralinos and charginos. This mass hierarchy substantially altered the decays of the sleptons, leading to final states containing SM bosons instead of the naively expected leptons. An additional feature of the R-sneutrinos is that they have also the potential to explain the observed dark matter relic density \cite{Asaka:2005cn,Gopalakrishna:2006kr,Arina:2007tm,Page:2007sh,Belanger:2010cd,Dumont:2012ee,DeRomeri:2012qd,Faber:2019mti}.

In this paper we first update the bounds on this class of models, taking into account the latest available analyses. Moreover, we will explore the reach of the upcoming LHC run for these scenarios. Here we will show that there are cases where the corresponding bound can reach about 300 GeV. This motivates us to investigate to which extent a prospective international linear collider (ILC), running at a center of mass energy of 1 TeV, will be able to discover such a scenario. Moreover, we will demonstrate that in spite of the drastically changed signatures, a rather precise mass measurement will still be possible at the ILC.

The paper is structured as follows: in next section we present some main features of 
the model and discuss the parameter regions of interest. In section~\ref{sec:LHC} we present our update of the bounds on the slepton masses taking into account the recent analyses. In Sec.~\ref{sec:sensitivity} we discuss the sensitivity of the ILC for such scenarios, and in the subsequent Sec.~\ref{sec:mass_reconstruction} we adapt a method for the reconstruction of chargino and neutralino masses to our case. We demonstrate that a mass measurement with a precision of 1-3 percent should be possible. Moreover, we collect in the appendix
the information on the software used in the various stages of this investigation.

\section{Scenarios of Interest}
\label{sec:model}

For this model, we start with the MSSM superpotential, and add three singlet superfields $\hat\nu_{R}$, such that R-parity is conserved. The superpotential is:
\begin{align}
\mathcal{W}_{{\rm eff}} = \mathcal{W}_{\rm MSSM}
+ \frac{1}{2} (M_R)_{ij}\,\hat{\nu}_{Ri}\,\hat{\nu}_{Rj}
+ (Y_\nu)_{ij}\,\widehat{L}_i \cdot \widehat{H}_u\, \hat{\nu}_{Rj}
\end{align}
The MSSM symmetries allow for the following soft SUSY-breaking terms:
\begin{equation}
\mathcal{V}^{soft} =\mathcal{V}_{\rm MSSM}^{soft}
  + (m^2_{\tilde\nu_R})_{ij}\tilde{\nu}^*_{Ri}\tilde{\nu}_{Rj}
  + \bigg(\frac{1}{2}(B_{\tilde\nu})_{ij}\tilde{\nu}_{Ri}\tilde{\nu}_{Rj} + (T_\nu)_{ij}\,\tilde{L}_i \cdot H_u\, \tilde{\nu}_{Rj} + \text{H.c.} \bigg)
\end{equation}

Based on naturalness arguments~\cite{Papucci:2011wy,Hall:2011aa}, we assume the hierarchy $\mu\ll M_{1,2,3}$, such that the lightest electroweakinos (neutralinos and chargino) have a dominant Higgsino component. We also assume that at least one slepton family is lighter than the electroweakinos. Squarks will be assumed to be decoupled. 

We take neutrino oscillation parameters compatible with the results in~\cite{deSalas:2017kay,Esteban:2018azc}, with a normal hierarchy. We assume that heavy neutrino masses are low enough such that the $\tilde\nu_R$ can be the lightest supersymmetric partners (LSP). For definiteness, we take two heavy neutrinos to have 20 GeV masses, with the third one being much lighter\footnote{The lightest heavy neutrino could contribute to the dark matter relic density, but requires a resonant production mechanism, such as in~\cite{Shi:1998km}.}. With more than one heavy neutrino, the several entries of the corresponding Yukawa matrix can be enhanced with respect to the naive seesaw expectation~\cite{Casas:2001sr,Donini:2012tt}. We set:
\begin{subequations}
\label{eq:YukawasSimple}
\begin{eqnarray}
 (Y_\nu)_{a 4} &=& (U_{\rm PMNS})_{a1}^*\sqrt{\frac{2m_1 M_4}{v_u^2}}~, \\
 (Y_\nu)_{a 5} &=& i\, z_{56} \,Z_a^*\sqrt{\frac{2m_3 M_5}{v_u^2}}\cosh\gamma_{56}\,e^{-i\,z_{56}\,\rho_{56}}~, \\
 (Y_\nu)_{a 6} &=& -Z_a^*\sqrt{\frac{2m_3 M_6}{v_u^2}}\cosh\gamma_{56}\,e^{- i\,z_{56}\,\rho_{56}}~.
\end{eqnarray}
\end{subequations}
Here, $m_i$ ($M_i$) are the masses of the light (heavy) neutrinos. The parameters $\rho_{56}$ and $\gamma_{56}$ are the real and imaginary components of a complex mixing angle within the full $6\times6$ neutrino mixing matrix, with $z_{56}$ the sign of $\gamma_{56}$. The $Z_a$ factors~\cite{Gago:2015vma,Jones-Perez:2019plk}, with $a=e,\,\mu,\,\tau$, depend on the PMNS mixing matrix $U_{\rm PMNS}$ and ratios of light neutrino masses. With the exception of $Z_e$, which is slightly suppressed, they are all of $\ord{1}$.

The $\gamma_{56}$ parameter is responsible for enhancing the Yukawas, and we take it large enough such that the NLSP is not long-lived. By setting $\gamma_{56}=8$, we obtain $|Y_{a5}|=|Y_{a6}|\sim\ord{10^{-4}}$. In principle, such an enhancement could imply correlations between the $\nu_R$ and $\tilde\nu_R$ phenomenology, which we do not pursue here. We check that our setup respects constraints from lepton flavor violation (LFV), neutrinoless double beta decay and direct searches~\cite{Atre:2009rg,Blennow:2010th,Alonso:2012ji,LopezPavon:2012zg,Gago:2015vma,Deppisch:2015qwa,Abada:2018sfh,Hernandez:2018cgc}.

For simplicity, we take $B_{\tilde \nu}=T_\nu=0$. In addition, to avoid issues with SUSY contributions to LFV, we take diagonal $m^2_{\tilde L}$, $m^2_{\tilde e_R}$, and $m^2_{\tilde \nu_R}$ soft masses. As was done in~\cite{Cerna-Velazco:2017cmn}, there is no need to separate the sneutrino into scalar and pseudoscalar components. We can safely assume that three $\tilde\nu_i$ states shall be dominantly $\tilde\nu_L$ (L-sneutrinos, $\tilde\nu_{L e}$, $\tilde\nu_{L\mu}$, $\tilde\nu_{L\tau}$), and other three states shall be dominantly $\tilde\nu_R$ (R-sneutrinos, $\tilde\nu_{1,2,3}$). Unless otherwise noted, we fix $\mu=500$~GeV, $\tan\beta=6$, and the soft $m_{\tilde\nu_R}=100$~GeV.

We will now report typical slepton mass patterns and branching ratios. First, for $m_{\tilde L}=m_{\tilde E}$, all charged sleptons will always be heavier than L-sneutrinos, with a possible exception for the lightest stau. This is due to the lepton mass and D-term contributions. In particular, the mass splittings due to D-terms are:
\begin{align}
  (m_{\tilde\ell_L}-m_{\tilde\nu_L})_D\approx\frac{(\sin^2\theta_W-1)m^2_Z\cos2\beta}{2m_{\tilde L}} & & 
  (m_{\tilde\ell_R}-m_{\tilde\nu_L})_D\approx\frac{(-\sin^2\theta_W-\tfrac{1}{2})m^2_Z\cos2\beta}{2m_{\tilde R}}
\end{align}
which are larger than zero for $\tan\beta>1$. For $m_{\tilde L}=m_{\tilde E}=300$\,GeV and $\tan\beta=6$, we find $(m_{\tilde\ell_L}-m_{\tilde\nu_L})_D\approx 10$\,GeV and $(m_{\tilde\ell_R}-m_{\tilde\nu_L})_D\approx 9$\,GeV.

This means that charged sleptons can decay into an L-sneutrino and fermions, via an off-shell $W$: $\tilde\ell\to\tilde\nu_{L\ell}\, W^{*}$. The subsequent decay of the $\tilde\nu_L$ would lead to a cascade. In this channel, given the relatively small mass splitting, the fermions are very soft. This will be true for both L- and R-sleptons, the latter decaying through their small L-R admixture.

\begin{figure}[tph]
	\centering
\includegraphics[width=0.45\textwidth]{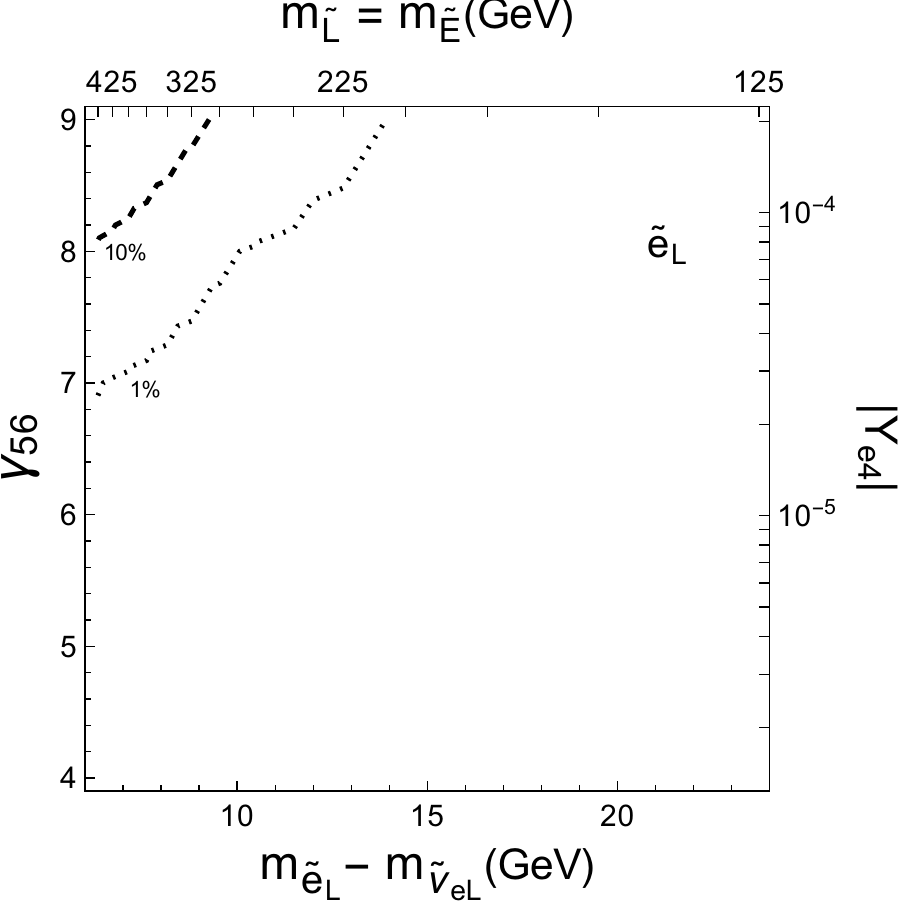} \hfill
\includegraphics[width=0.45\textwidth]{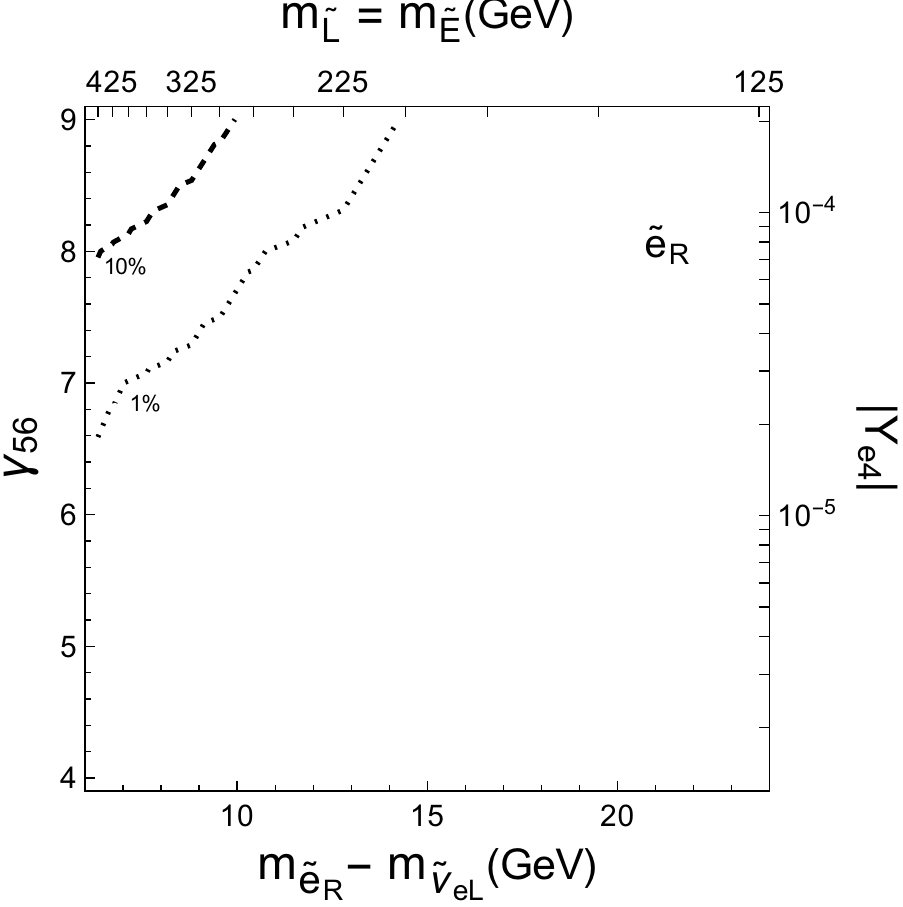} \\
\vspace{1cm}
\includegraphics[width=0.45\textwidth]{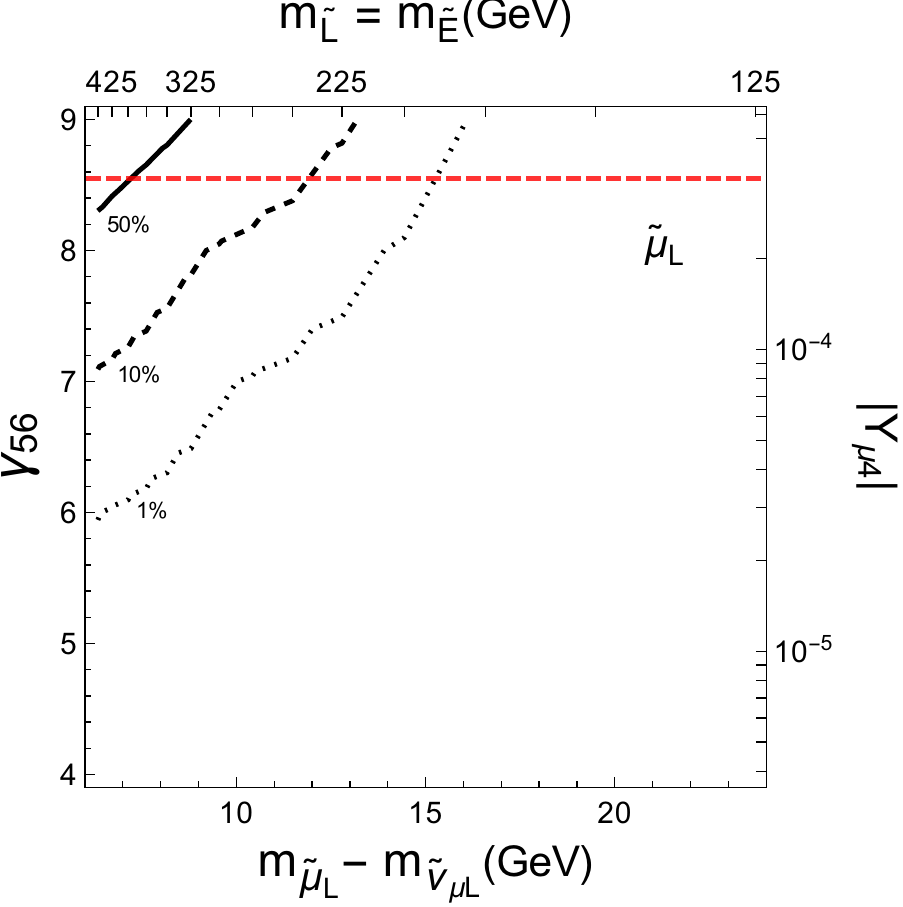} \hfill
\includegraphics[width=0.45\textwidth]{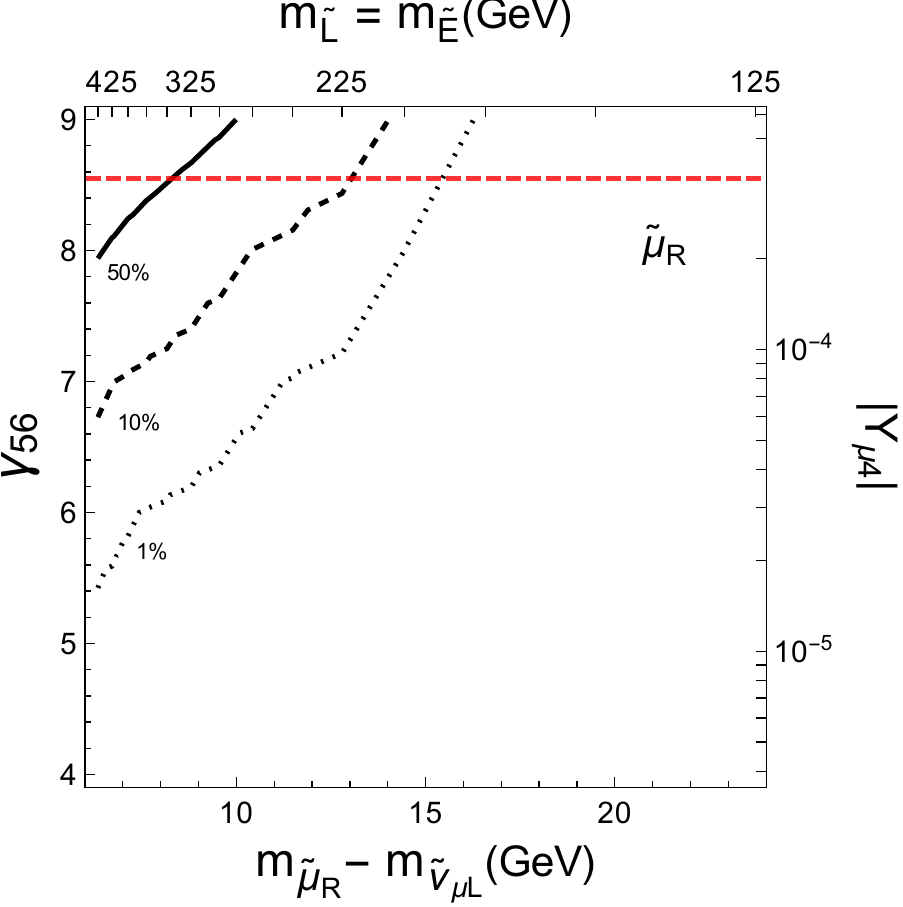}
\caption{Branching ratios of first two generations of charged sleptons into $\tilde\nu_R\, W$ final states, with $m_{\tilde\nu_R}=100$~GeV. We show branching ratios for $\tilde e_L$, $\tilde e_R$, $\tilde\mu_L$, and $\tilde\mu_R$ on the top left, top right, bottom left, and bottom right panels, respectively. Solid, dashed and dotted contours refer to $50\%$, $10\%$, and $1\%$. The region above the horizontal red line is excluded by direct searches of heavy neutrinos~\cite{Sirunyan:2018mtv}, which can bound specific Yukawa couplings.}
\label{fig:BRonshell}
\end{figure}
Another possible decay mode involves an on-shell $W$: $\tilde\ell\to\tilde\nu_R\, W$. In this channel, since the R-sneutrino couples to the vector boson through L-R mixing, and considering that we are taking $T_\nu=0$, the partial width is proportional to $Y_\nu$. The Yukawa suppression competes with the phase space suppression in the off-shell channel and, for small enough splitting or large enough $Y_\nu$ enhancement, the on-shell channel becomes dominant. The interplay between the $Y_\nu$ and phase space suppressions is shown in Fig.~\ref{fig:BRonshell}, where the on-shell branching ratio is displayed for the first two generations.

The third generation has a different decay pattern. For the lightest stau, the negative contribution from L-R mixing can bring $m_{\tilde\tau_1}$ much closer to the L-sneutrino mass. This negative contribution is approximately:
\begin{equation}
  \label{eq:splitLR}
  (m_{\tilde\tau_1}-m_{\tilde\ell})_{LR}\sim-\frac{m_\tau\, \mu\tan\beta}{2m_{\tilde L}}
\end{equation}
We find that for $\mu=500\,$GeV, $m_{\tilde L}=300\,$GeV and $\tan\beta=6$, the $\tilde\tau_1-\tilde\nu_{L\tau}$ mass splitting is around 1~GeV. The stau can even become lighter than the L-sneutrino for the same $\tan\beta$ but $\mu\gtrsim570\,$GeV, or for the same $\mu$ and $\tan\beta\gtrsim7$, a scenario we do not pursue in this work.

\begin{figure}[t]
	\centering
\includegraphics[width=0.49\textwidth]{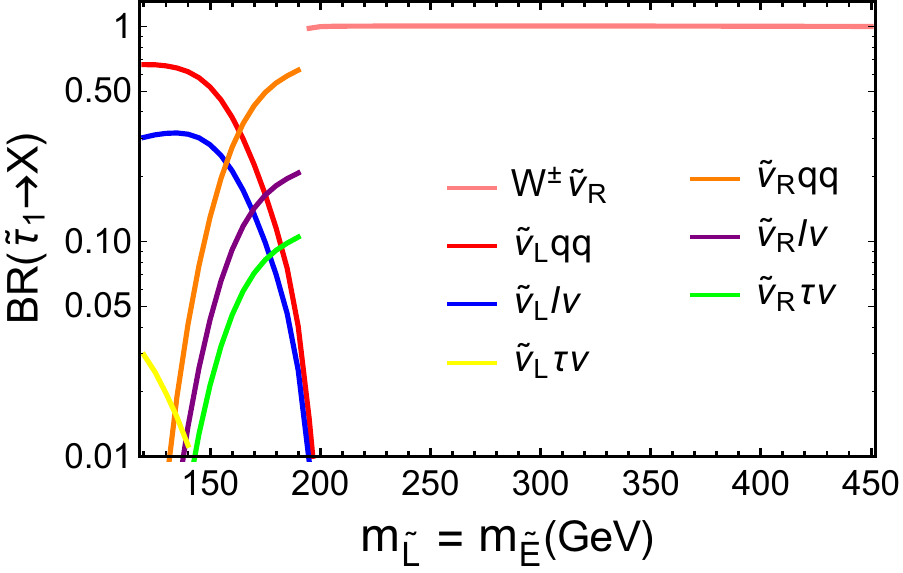} \hfill
\includegraphics[width=0.49\textwidth]{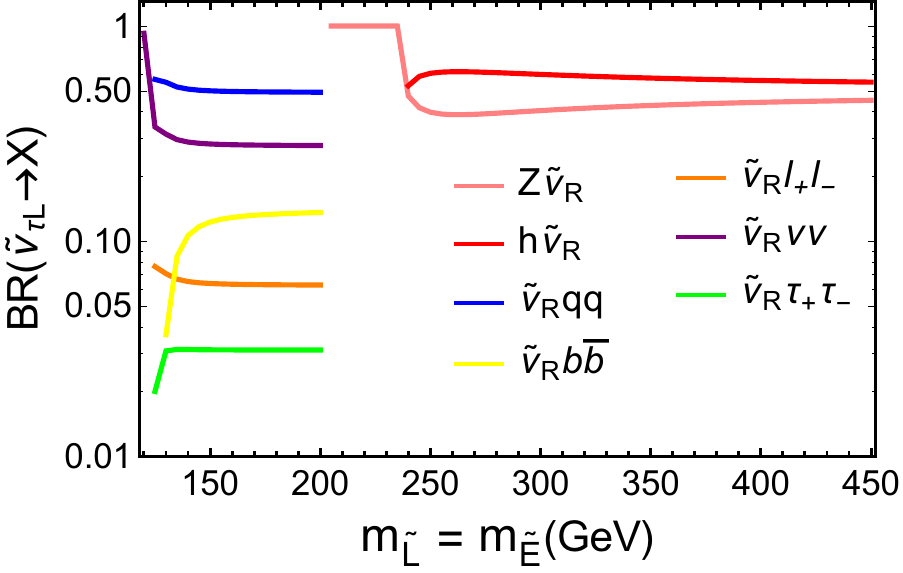}	
\caption{Branching ratios of $\tilde\tau_1$ (left) and $\tilde\nu_{L\tau}$ (right), as a function of their soft mass parameter. We fix $\gamma_{56}=8$ and $m_{\tilde\nu_R}=100$~GeV.}
\label{fig:BR}
\end{figure}
Given the smaller mass splitting, the dominant decay channel of the $\tilde\tau_1$ will usually be into a $\tilde\nu_R$ and a $W$, as long as it is phase-space allowed. This is shown on the left panel of Fig.~\ref{fig:BR}. For small masses the phase space suppression is not so strong, so the $\tilde\tau_1$ decays as the other charged sleptons: into a $\tilde\nu_{L\tau}$ and an off-shell $W$.

On the other hand, the heaviest stau will have an increase in mass opposite to that in Eq.~(\ref{eq:splitLR}). Given the larger phase space, the $\tilde\tau_2$ have mostly decays into a $\tilde\nu_{L\tau}$ and an off-shell $W$, with a small chance for decay into a $\tilde\tau_1$ and an off-shell $Z$.

Finally, the decay modes of the $\tilde\nu_{L\tau}$ are shown on the right panel of Fig.~\ref{fig:BR}. On the large mass limit, the $\tilde\nu_{L\tau}$ decays into a $\tilde\nu_R$ and either a $Z$ or $h$ boson, both with around $50\%$ probability. For lower masses, the bosons go off-shell. Notice there is a small window where the $\tilde\nu_{L\tau}$ will decay through the on-shell $Z$ channel with $100\%$ probability. All L-sneutrinos decay in the same way.

In the following, we will explore the reach of both LHC and ILC when searching for these sleptons. We focus on three different scenarios, on all of them fixing $\gamma_{56}=8$:
\begin{itemize}
    \item In scenario SE, we assume that the only light MSSM sleptons are the $\tilde e_L,\,\tilde e_R$ and $\tilde\nu_{eL}$. This kind of situation can be justified by specific flavor symmetries~\cite{Jones-Perez:2013uma}. Here, L-R mixing is unimportant.
    
    \item In scenario ST, we take the $\tilde\tau_1,\,\tilde\tau_2$, and $\tilde\nu_{L\tau}$ as the light MSSM sleptons, which is common in split-family SUSY~\cite{Cohen:1996vb,Craig:2011yk,Delgado:2011kr,Larsen:2012rq,Craig:2012hc,Blankenburg:2012nx}. An important observation is that the sleptons decay in the same way as light electroweakinos~\cite{Battaglia:2006bv}. Thus, we expect searches targeting such models to be sensitive to this scenario. 
    
    \item In scenario DEG, we consider the situation where all MSSM sleptons share the same soft masses, $m_{\tilde L}=m_{\tilde E}$.
    
\end{itemize}

\section{Update to Slepton Searches at the LHC}
\label{sec:LHC}

At the LHC, the most important slepton production processes are:
\begin{align}
p\,p \xrightarrow{} \tilde{\ell}_L\,  \tilde{\nu}_L, & &
p\,p \xrightarrow{} \tilde{\tau}_{1,2}\, \tilde{\nu}_{\tau}
\end{align}
where $\tilde\ell$ refers to any charged slepton except the stau. The produced sleptons will decay with the branching ratios reported in the previous section, meaning we are interested in the following final states:
\begin{align}
W^{(*)}+(Z/h)^{(*)}+p_T^{\rm miss} & & 2(Z/h)^{(*)}+p_T^{\rm miss}+{\rm soft\,fermions}
\end{align}
From our analysis, we expect events in scenario SE to be characterized by $ZZ$, $hh$ and $hZ$ final states, with one of the bosons coming from a cascade decay of a charged slepton accompanied by soft jets.  Of course, the bosons will be on-shell provided the $\tilde\nu_L$ are heavy enough. For lighter sneutrinos, the off-shell $Z$ bosons would lead to high $p_T$ fermions. For scenario ST, we expect a sufficiently heavy $\tilde\tau_1$ to decay into an on-shell $W$ instead, giving $WZ$ and $Wh$ final states. The DEG scenario will naturally include a combination of both cases.

An early scan of the parameter space of the DEG scenario was carried out in~\cite{Cerna-Velazco:2017cmn}, finding that the region with $m_{\tilde L}<\mu$ was very poorly constrained, namely ruling out masses $m_{\tilde L}<150\,$GeV. At the time, this was attributed to the lack of searches targeting SUSY decays involving $W$, $Z$ and $h$ bosons, typical of our scenario of interest. Thus, now that new searches have been carried out by both ATLAS and CMS, we present an update of the relevant exclusion region in~\cite{Cerna-Velazco:2017cmn}, including also the SE and ST scenarios. As before, we also include channels with smaller cross section, $pp\to\tilde\ell\,\tilde\ell^*$ and $pp\to\tilde\nu_L\tilde\nu_L^*$.

The most relevant analysis for our model is the CMS search for two or more leptons and missing energy, at 13 TeV and $35.9\,{\rm fb}^{-1}$~\cite{Sirunyan:2017lae}. Their analysis considers a large number of signal regions, where leptons pairs can have an invariant mass lower, larger or consistent with the $Z$ boson mass. They interpret their results in simplified electroweakino models, and rule out masses up to around $450$~GeV in the $\tilde\chi^+_1\tilde\chi^0_2\to W\,Z\,\tilde\chi^0_1\,\tilde\chi^0_1$ channel (for $m_{\rm LSP}$ lower than about 150~GeV), and around $175$~GeV for $\tilde\chi^+_1\tilde\chi^0_2\to W\,h\,\tilde\chi^0_1\,\tilde\chi^0_1$ ($m_{\rm LSP}\lesssim25$~GeV). They also interpret their search in gauge-mediated SUSY breaking, ruling out masses below $\sim450$~GeV in the $\tilde\chi\,\tilde\chi\to Z\,Z\,\tilde G\,\tilde G$ channel (for a gravitino $\tilde G$ mass of 1~GeV), assuming the electroweakinos decay to $Z\,\tilde G$ states with $100\%$ probability.

We have found no other analysis capable of excluding any region of the parameter space better than~\cite{Sirunyan:2017lae}. Nevertheless, the ATLAS search for exactly two soft leptons and missing energy in~\cite{Aaboud:2017leg}, targeting models with compressed spectra, is the most sensitive search for the upper left border of the parameter space, where $m_{\tilde L}\sim m_{\tilde\nu_R}$.

\begin{figure}[t]
\centering
\includegraphics[width=0.48\textwidth]{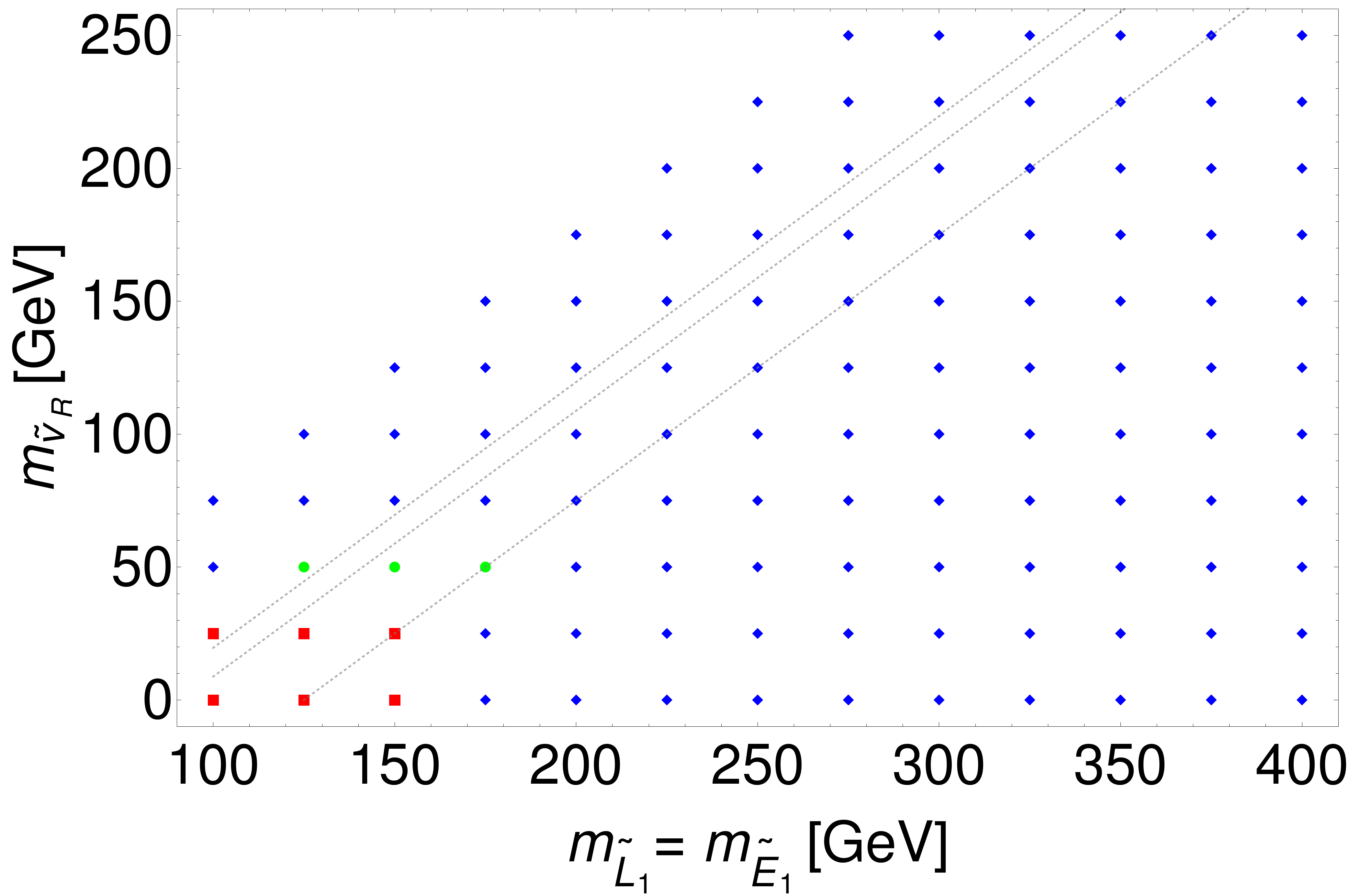}\quad 
\includegraphics[width=0.48\textwidth]{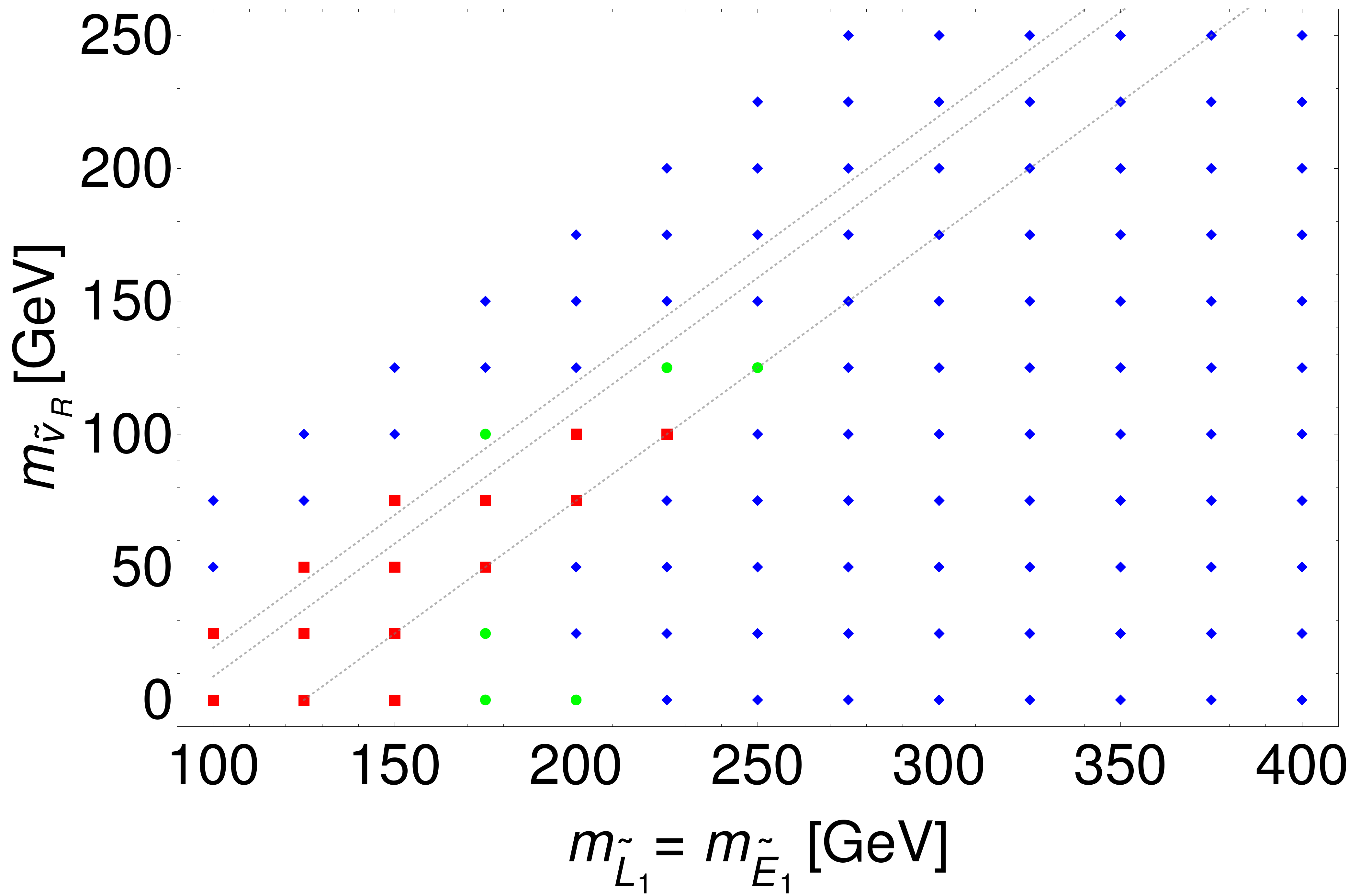}
\caption{Currently excluded points (left) and expectations for 300 fb$^{-1}$ (right), for scenario SE. Red (square) points are excluded, blue (diamond) points are allowed, and green (round) points are considered ambiguous due to theoretical and experimental uncertainties. Gray lines, from left to right, indicate $m_{\tilde L 1}=m_{\tilde\nu_R}+m_W$, $m_{\tilde L 1}=m_{\tilde\nu_R}+m_Z$, and $m_{\tilde L 1}=m_{\tilde\nu_R}+m_h$.}
\label{fig:excE}
\end{figure}

\begin{figure}[t]
\centering
\includegraphics[width=0.48\textwidth]{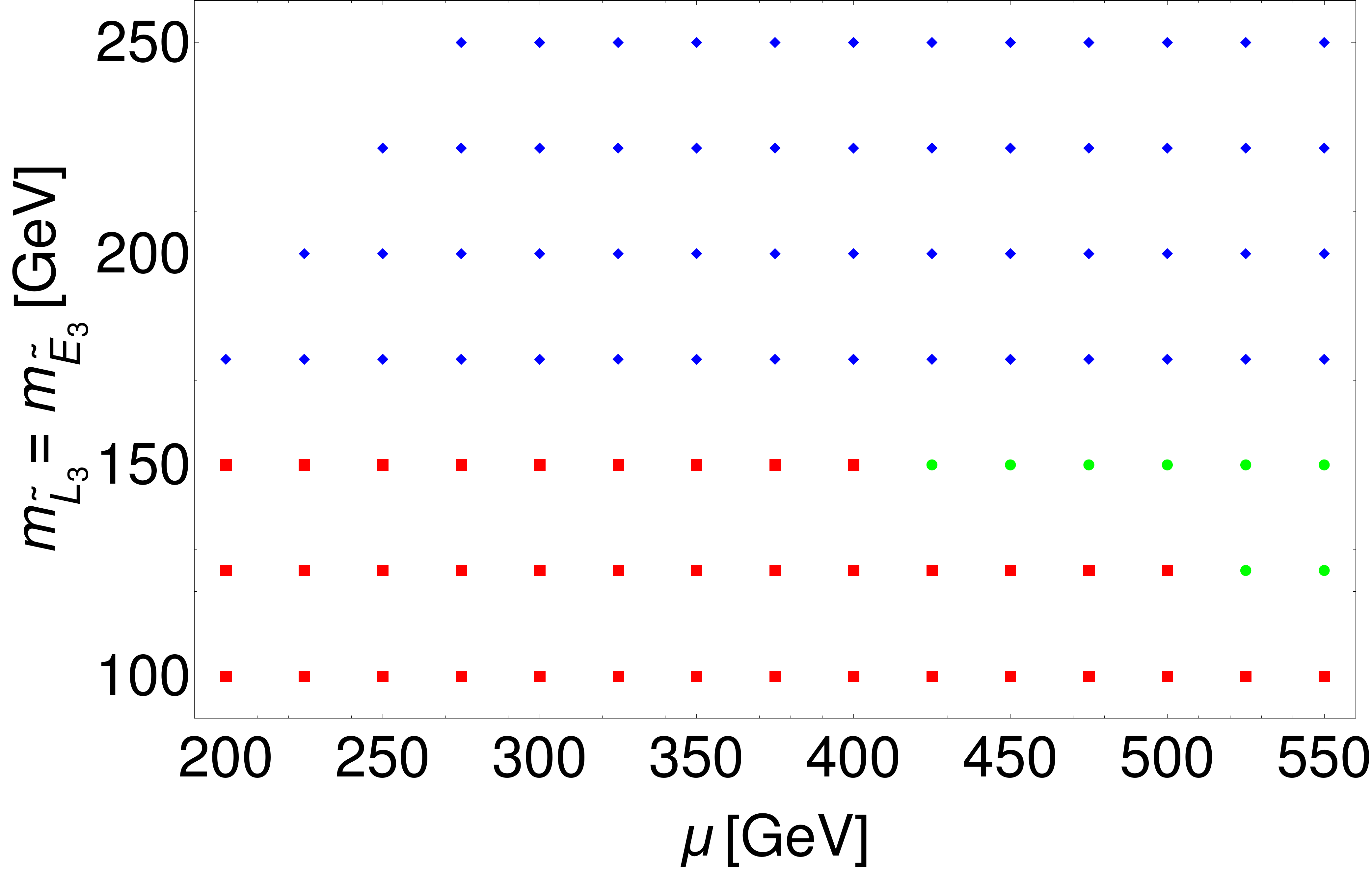}\quad 
\includegraphics[width=0.48\textwidth]{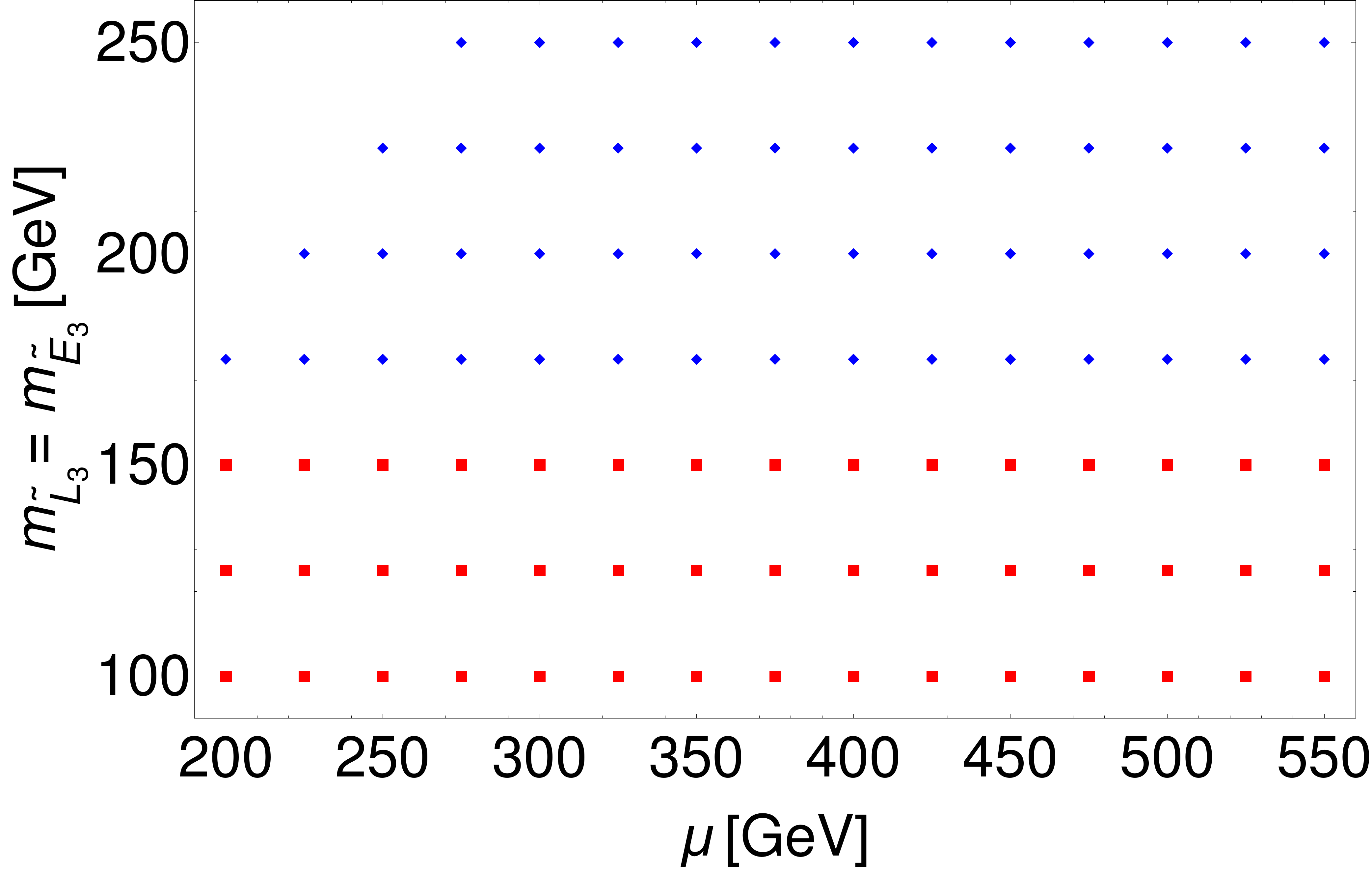}
\caption{As Fig.~\ref{fig:excE}, for scenario ST. Notice that here we vary $\mu$, setting $m_{\tilde\nu_R}=0$~GeV.}
\label{fig:excT}
\end{figure}

\begin{figure}[t]
\centering
\includegraphics[width=0.48\textwidth]{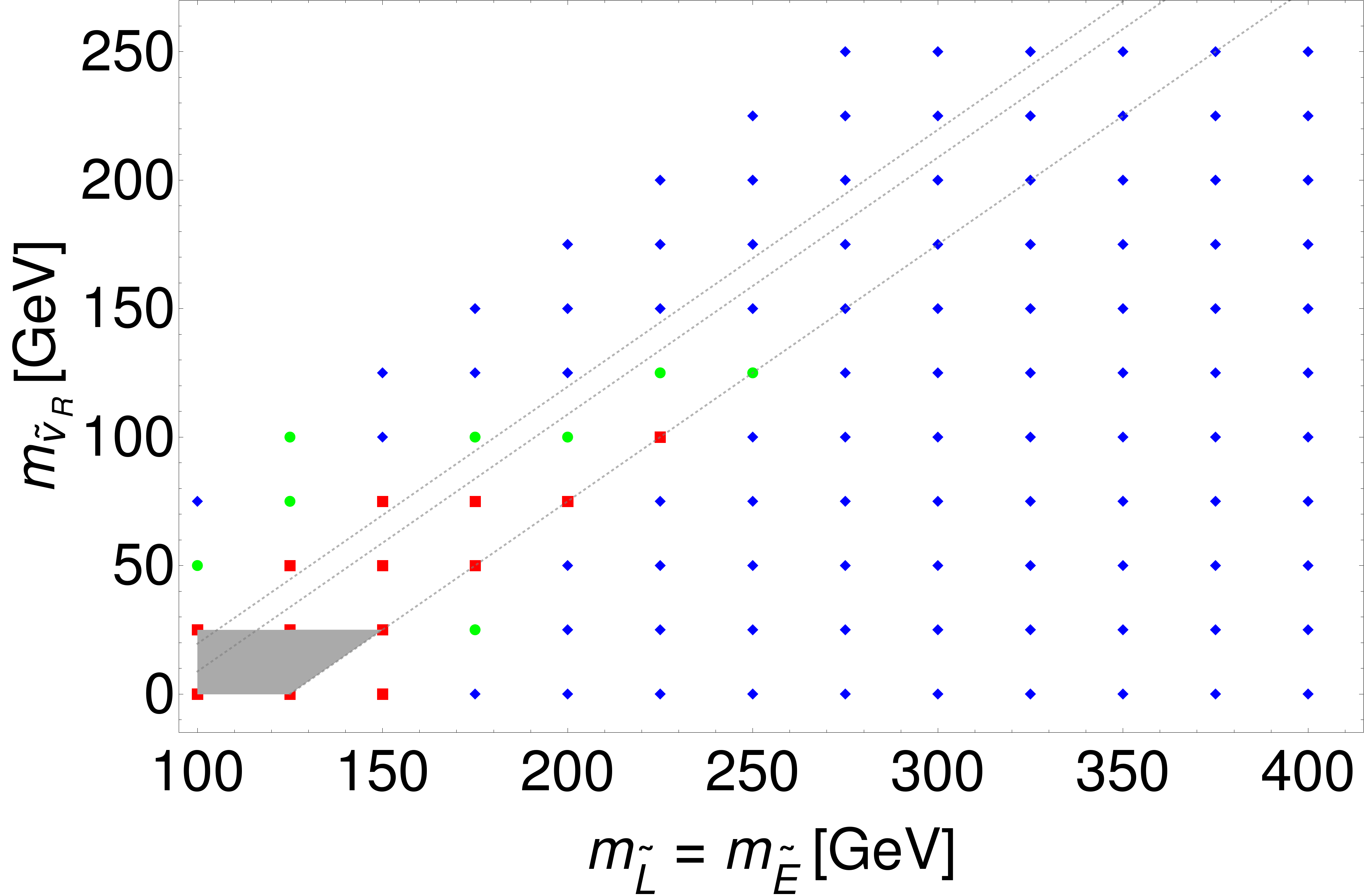}\quad 
\includegraphics[width=0.48\textwidth]{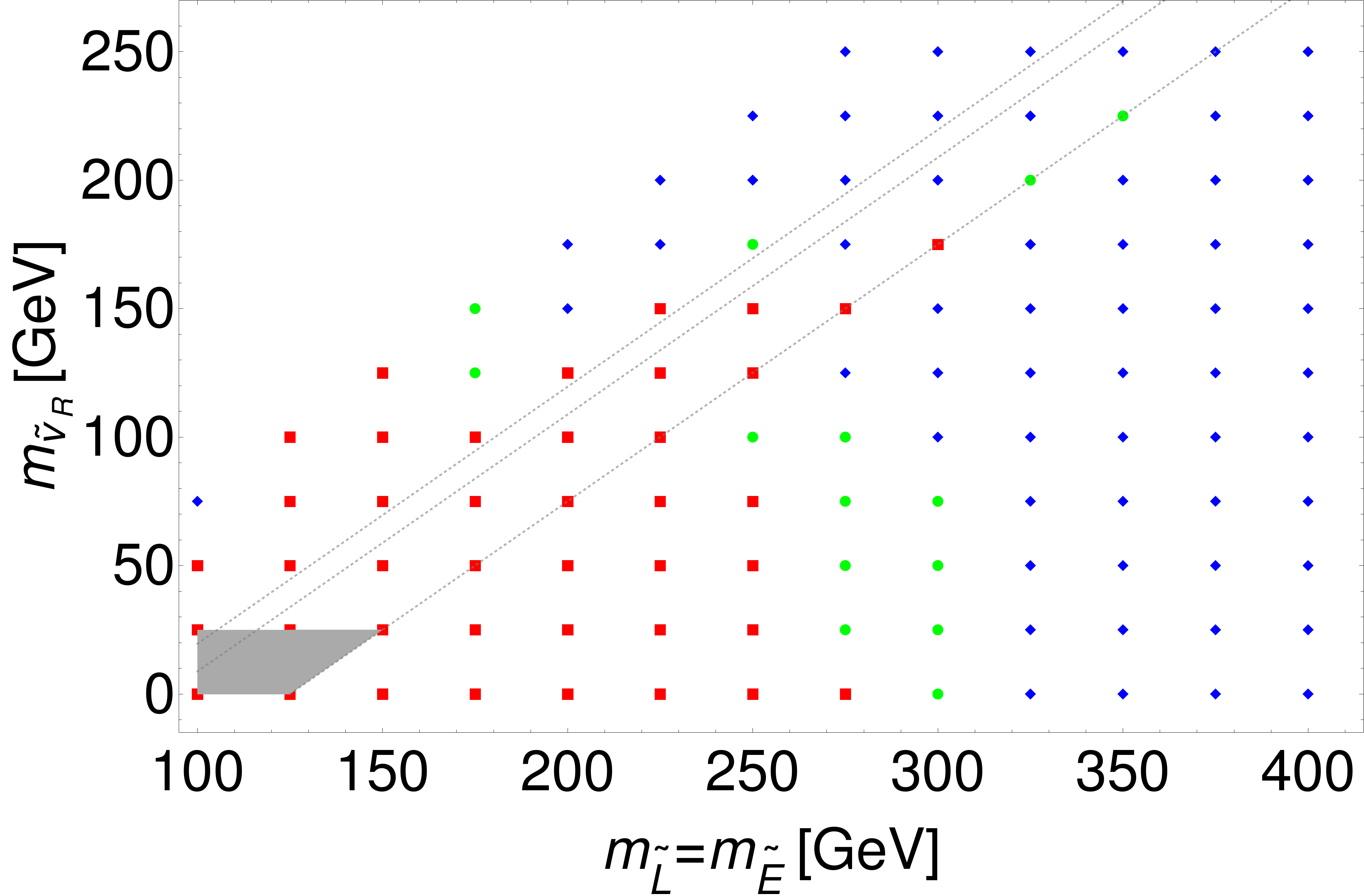}
\caption{As Fig.~\ref{fig:excE}, for scenario DEG. The shaded region indicates the constraints obtained in Fig. 8 of~\cite{Cerna-Velazco:2017cmn}.}
\label{fig:excD}
\end{figure}

The results of our scans are show in in Figs.~\ref{fig:excE}-\ref{fig:excD}. We show constraints from the currently available data, and the expectations for an increase of luminosity up to 300~fb$^{-1}$. On all plots, red (square) points are excluded and blue (diamond) points are allowed. We also define green (round) points as ambiguous, following the suggestion in~\cite{Drees:2015aeo}, due to theoretical uncertainties, such as the choice of parton distribution function, and the fact that experimental searches are not tailored to this specific model.

The exclusion regions for scenario SE are shown in Fig.~\ref{fig:excE}. We see that the reach of the search is currently very poor, with no bounds for $m_{\tilde\nu_R}\gtrsim50$~GeV. If lighter, selectrons are restricted to be heavier than around 125-150 GeV. Fortunately, this will improve for higher luminosity, where the search can exclude slepton masses up to 225~GeV. Nevertheless, having a point excluded or not depends entirely on the decay products. We find a very strong drop in sensitivity when $m_{\tilde L 1}\gtrsim m_{\tilde\nu_R}+m_h$. This is due to the opening of the $\tilde\nu_{Le}\to h\,\tilde\nu_R$ channel, happening with a 50\% probability, with the Higgs boson in turn having a very small branching ratio into the leptonic final states targeted by~\cite{Sirunyan:2017lae}. The sensitivity is also lost for $m_{\tilde L 1}\ll m_{\tilde\nu_R}+m_Z$, since then the final state leptons from the virtual $Z$ turn out too soft to be picked up by the detector.

For the ST scenario, we fixed $m_{\tilde\nu_R}=0$~GeV and explored the role of $\mu$ in the $\tilde\tau_1$ exclusion. This is motivated by the effect of L-R mixing on the physical stau mass, and thus on its branching ratios. Notice that even though the $\tilde\nu_R$ has a vanishing soft mass, its real mass is still around the heavy neutrino mass of 20 GeV. On our scan, we again found zero sensitivity when the $\tilde\nu_{L\tau}\to h\,\tilde\nu_R$ channel opened, at around 150 GeV. This would not improve for higher luminosity. Notice that this means there are no restrictions on $\mu$ coming from slepton production\footnote{We will address constraints on $\mu$ from other sectors of the model in a future work.} if these are heavier than 150 GeV. For lighter staus, $\mu$ is currently bounded to values above 425-550 GeV. Sensitivity is lost for larger values due to the $\tilde\tau_1$ becoming too light to produce an on-shell $W$, with the decay product thus becoming too soft. As shown on the right panel, these points can be probed by adding data. We do not analyze larger values of $\mu$, as this leads to the $\tilde\tau_1$ being much lighter than $\tilde\nu_L$, which changes the phenomenology of the model.

Finally, we report results for the DEG scenario in Fig.~\ref{fig:excD}, again fixing $\mu=500$~GeV and varying $m_{\tilde\nu_R}$. Here, we get twice as many events as in the SE scenario, due to selectron and smuon decays having an almost identical phenomenology. This duplication of events allows the future probing of the $m_{\tilde L}\gtrsim m_{\tilde\nu_R}+m_h$ region, up to almost $m_{\tilde L}\sim300$~GeV. The addition of $\tilde\tau$ data is also useful for constraining the low $m_{\tilde L}$ region, being able to exclude $m_{\tilde L}\lesssim150$~GeV when $m_{\tilde L}\sim m_{\tilde\nu_R}$. By comparing with the shaded region we can appreciate the improvement brought by~\cite{Sirunyan:2017lae,Aaboud:2017leg}.

To summarize, single slepton families are currently constrained to being heavier than about 150~GeV, for light R-sneutrinos. In the future, these constraints can be somewhat raised to masses of order 200~GeV, with decreased sensitivity in the $m_{\tilde L}\gtrsim m_{\tilde\nu_R}+m_h$ and $m_{\tilde L}\sim m_{\tilde\nu_R}$ regions. For degenerate soft masses, the current situation is better than the one reported in~\cite{Cerna-Velazco:2017cmn}, with current exclusions reaching about 225~GeV, and future bounds expected to increase up to 300~GeV. Nevertheless, the diminished sensitivity in the $m_{\tilde L}\gtrsim m_{\tilde\nu_R}+m_h$ region reported for the single slepton case is also present here. For $m_{\tilde L}\sim m_{\tilde\nu_R}$, the coverage improves significantly with respect to the single slepton scenarios.

It is thus clear that with current searches the LHC shall not be able to probe the entire parameter space shown in Figs.~\ref{fig:excE}-\ref{fig:excD}. This motivates studies at future colliders, such as the HL-LHC and ILC. In the following, we turn toward the latter.

\section{Sensitivity at ILC}
\label{sec:sensitivity}

The International Linear Collider (ILC) is a proposed experiment, most likely located in the Japanese highlands, that will generate $e^+ e^-$ collisions with an initial energy of 250~GeV and 2~ab$^{-1}$ of integrated luminosity~\cite{Barklow:2015tja,Aihara:2019gcq}. From the exclusion regions presented in the previous section, we expect that for this energy the ILC will make no improvements with respect to the LHC reach. However, the ILC Technical Design Report also includes details on possible upgrades to the centre of mass energy and luminosity, of up to 1~TeV and 8~ab$^{-1}$, respectively~\cite{Baer:2013cma,Barklow:2015tja,Moortgat-Picka:2015yla}. In the following, we concentrate on this upgrade and evaluate the sensitivity to our signal.

Given our findings regarding the expected reach of the LHC, and following the branching ratios reported in Figs.~\ref{fig:BRonshell} and~\ref{fig:BR}, the primary channels of interest are:
\begin{align}
\label{eq:prodprocess1}
e^- e^+ \xrightarrow{} \tilde{\ell}^-  \tilde{\ell}^+, & \quad
\tilde{\ell}^\pm \xrightarrow{} \tilde{\nu}_L\, ff' \\
e^- e^+ \xrightarrow{} \tilde{\nu}_L \tilde{\nu}_L, & \quad
\tilde{\nu}_L \xrightarrow{} \tilde{\nu}_R\, Z/h \\
\label{eq:prodprocess3}
e^- e^+ \xrightarrow{} \tilde{\tau}_{1}^-\, \tilde{\tau}_{1}^+, & \quad 
\tilde{\tau}_1^\pm \xrightarrow{} \tilde{\nu}_R\, W^\pm
\end{align}
where $\tilde{\ell}$ denotes every slepton other than the lightest $\tilde{\tau}$. In addition, $\tilde\tau_2$ can also be produced, decaying similarly to $\tilde\ell$, but with an additional channel involving $\tilde\tau_1$ and soft fermions. Again, the $\tilde\nu_R$ are stable, so our final states always contain considerable missing energy and two on-shell SM bosons.

At $\sqrt{s}=1\,$TeV the ILC is expected to run mainly in two different polarization configurations with equal amounts of data~\cite{Barklow:2015tja}. Type~\textbf{L} polarization ($e^-_Le^+_R$) is intended to study the Higgs boson properties and searching for signals from specific BSM scenarios, such as composite Higgs models. Type~\textbf{B} polarization ($e^-_Re^+_L$) is to be used for general searches of new physics, in particular Supersymmetry, as it greatly reduces the SM background. In the following, we seek to confirm if, within our scenario, type \textbf{B} polarization is still the most convenient. As indicated in~\cite{Baer:2013cma}, we set the electrons (positrons) to be 80\% (20\%) polarized.

\begin{figure}[t]
\centering
\includegraphics[width=0.49\textwidth]{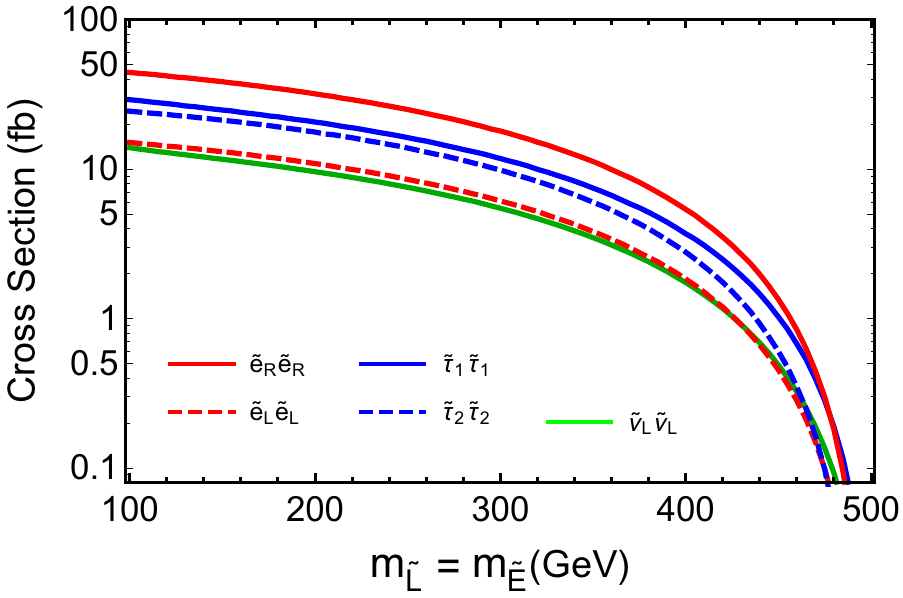}\hfill
\includegraphics[width=0.49\textwidth]{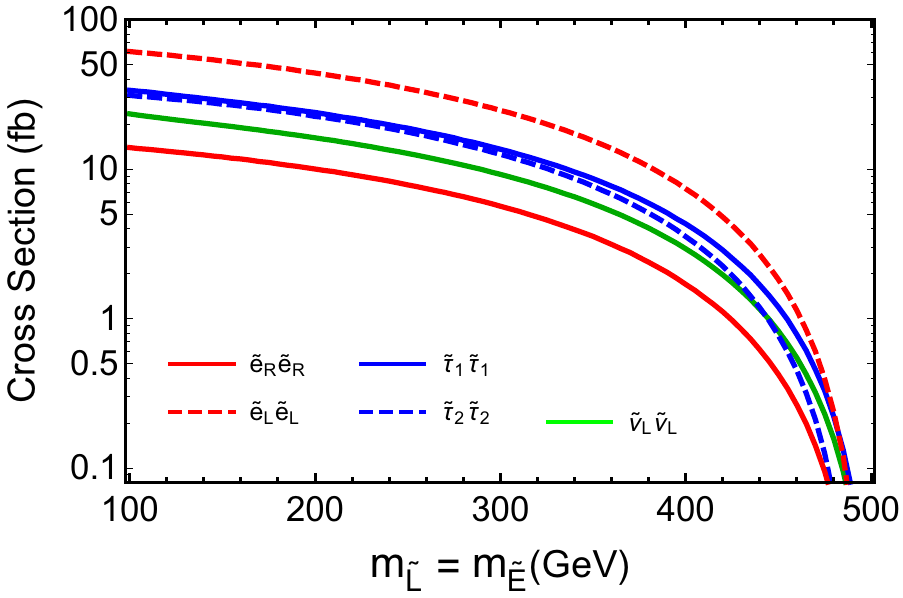}
\caption{Production cross section for $\sqrt s=1$~TeV. Type \textbf{B} (\textbf{L}) polarization is shown on the left (right) panel. Cross sections for L-sneutrinos, selectrons and staus are given in green, red, and blue lines, respectively.
For the $\tilde\tau$, solid (dashed) line indicate the lightest (heaviest) stau. For the $\tilde e$, solid (dashed) line refer to $\tilde e_R$ ($\tilde e_L$).}
\label{fig:cross}
\end{figure}
The cross section for slepton pair production is shown in Fig.~\ref{fig:cross} as a function of the soft slepton masses $m_{\tilde{L}}, m_{\tilde{E}}$. We show results for both types of polarizations, and for the three scenarios of interest. Due to the large scale of $M_1$ and $M_2$, the gaugino mediated contribution to $e^+e^-\to\tilde e^+\tilde e^-$ is negligible. Therefore, the total cross section for a single slepton family can be taken independent of flavor to an excellent approximation, depending only on $m_{\tilde{L}}$ and $m_{\tilde{E}}$.

In order to probe the processes indicated in Eqs~(\ref{eq:prodprocess1})-(\ref{eq:prodprocess3}), we impose the following cuts, adapted from~\cite{Suehara:2009bj}:
\begin{itemize}

\item Missing transverse momentum $p_T^{\rm miss}> 50$ GeV. The cut is kept relatively small since $m_{\tilde{\nu}_R}$ can be of the order of a few GeV.

\item Exactly four jets or b-jets with $p_T> 20$ GeV. This reduces considerably the six-jet backgrounds from SM processes, such as $ZWW$ production. In addition, the cut will improve slepton mass reconstruction, and reduce SUSY backgrounds involving jets from $\tilde\ell^\pm\to\tilde\nu_{L} f f'$ decays.

\item Two reconstructed SM bosons. The reconstruction requires finding two pairs of dijets with invariant masses $m_1$, $m_2$, such that they minimize:
\begin{equation}
f(m_1,\,m_2)=\dfrac{(m_1-{m_B}_1)^2+(m_2-{m_B}_2)^2}{\sigma^2}
\end{equation}
Here $\sigma=5$, and ${m_B}_i$ denote the masses of either $W^\pm$, $Z$ or Higgs boson. In general ${m_B}_1$, ${m_B}_2$ need not be equal. For Higgs boson reconstruction, the dijet must consist of b-jets. For $Z$ bosons, we accept either jet or b-jet pairs. The cut requires:
\begin{equation}
 f(m_1,\,m_2)<4
\end{equation}

\item No leptons with $p_T> 25$ GeV.  The reconstruction involves only jets, so this lepton veto is applied to reduce semi-leptonic backgrounds.

\item The angle between the beam direction and $p_T^{\rm miss}$ is constrained such that $|\cos(\theta_{\rm miss})|<0.99$, in order to reduce background from coplanar events.
    
\end{itemize}

\begin{table}[t]
\centering
\setlength{\tabcolsep}{1.25em}
{\begin{tabular}{| c | c | c|  c|  c|} 
\hline
 & \multicolumn{2}{|c|}{Type \textbf{B}} & \multicolumn{2}{|c|}{Type \textbf{L}} \\
\hline
$e^+e^-\xrightarrow{}$ & Events &  Efficiency ($\%$) & Events &    Efficiency ($\%$)\\
\hline
$W^+W^-$ & 13 & 0.005 & 90 & 0.003\\
\hline
$\nu\nu Z$ & 2 & 0.003 & 23 & 0.003\\
\hline
$t\bar{t}$ & 101 & 0.1 & 256 & 0.2\\
\hline
$ZZ$ & 36 & 0.06 & 75 & 0.05\\
\hline
$\nu\nu h$ & 1 & 0.003 & 11 & 0.005\\
\hline
$Zh$ & 52 & 0.7 & 81 & 0.6\\
\hline
$ZW^+W^-$ & 74 & 1 & 902 & 1 \\
\hline
$b\bar{b}b\bar{b}$ & 2 & 0.07 & 6 & 0.08\\
\hline
$\nu\nu W^+W^-$ & 88 & 4 & 1132 & 4\\
\hline
$t\bar{t}b\bar{b}$ & 1 & 0.08 & 2 & 0.08\\
\hline
$\nu\nu ZZ$ & 34 & 5 & 315 & 5\\
\hline
$hW^+W^-$ & 3 & 0.7 & 24 & 0.6\\
\hline
$ZZZ$ & 7 & 3 & 28 & 3\\
\hline
$hZZ$ & 1 & 1 & 5 & 2\\
\hline
$hhZ$ & 1 & 0.8 & 1 & 0.8\\
\hline
$\nu\nu hh$ & 1 & 2 & 2 & 2\\
\hline\hline
All background & 417 & 0.08 & 2950 & 0.06 \\
\hline \hline
All signal, scenario SE & 758 & 5 & 1019 & 5 \\
All signal, scenario ST & 922 & 6 & 1245 & 6 \\
All signal, scenario DEG & 2413 & 6 & 3232 & 6 \\
\hline
\end{tabular}}
\caption{Number of background events after cuts for various processes and for both types of polarization, considering a luminosity of 500~fb$^{-1}$.  The last rows includes the sum of events from all signal processes, including cascades, for comparison. The efficiency columns refer to the ratio between the number of events after the cuts over those initially generated.}
\label{table:cut_bkg_B}
\end{table}
In order to compare both polarizations, we report the number of events for the considered signal and background processes. Table \ref{table:cut_bkg_B} shows the background data in detail for type \textbf{B} and \textbf{L} polarization. For the signal, we take as a benchmark the point $m_{\tilde{L}}=m_{\tilde{E}}=300$~GeV, $m_{\tilde{\nu}_R}=100$ GeV. We find that for type \textbf{B} the most important background comes from $t\bar t$ production, followed by $\nu\,\nu\,W^+W^-$ and $Z\,W^+W^-$. Other important backgrounds are $Z\,h$, $Z\,Z$ and $\nu\,\nu\,Z\,Z$ production. In contrast, for type \textbf{L} polarization the dominating background comes from $\nu\,\nu\,W^+W^-$, followed closely by $Z\,W^+W^-$, and then by $\nu\,\nu\,Z\,Z$ and $t\bar t$.

We find the total background for type \textbf{L} polarization to be about one order of magnitude larger than that for type \textbf{B}. In contrast, the number of signal events for type \textbf{L} is not much larger than the one for type \textbf{B}. Therefore, we confirm that type \textbf{B} is still the most convenient, and consider only this polarization in the rest of this section.

\begin{table}[t]
\centering
\setlength{\tabcolsep}{1.5em}
{\begin{tabular}{| c | c | c|  c|} 
\hline
Scenario & SE & ST &  DEG\\
\hline
No cuts & 14713 & 14745& 44134\\
\hline
$p_T^{\rm miss}>50$ GeV & 12941 & 12997& 38850\\
\hline
Exactly four jets with $p_T>20$~GeV  & 4740 & 3770& 12948\\
\hline
Exactly two reconstructed SM bosons & 869 & 1092& 2901\\
\hline
$p^{\rm lepton}_T<25$ GeV &862 & 1084& 2878\\
\hline
$|\cos(\theta_{\rm miss})|<0.99$ & 758 & 922&2413 \\
\hline
\hline
Efficiency ($\%$) & 5.2 & 6.3& 5.5\\
\hline
\end{tabular}}
\caption{Cutflow for our three scenarios with type \textbf{B} polarization. We consider a luminosity of 500 fb$^{-1}$.}
\label{table:cutflow}
\end{table}
For our benchmark point, we show on Table \ref{table:cutflow} the corresponding cutflow in our three scenarios. One of the strongest cuts is the requirement of exactly four jets. This is due to the probability of having both $WW$ and $ZZ$ pairs decaying into jets being of about $50\%$. This is further affected by jet reconstruction and tagging efficiencies at the detector level. The next strongest constraint is the requirement of reconstructing two SM bosons.

We find that scenario ST has a slightly larger global efficiency. Regarding the 4-jet cut, this scenario has a further suppression of events due to the different decay modes of the $\tilde\tau_2$. Since the latter has a much larger mass, jets from $\tilde\tau_2\to\tilde\nu_{L\tau}\, q\, q'$ and $\tilde\tau_2\to\tilde\tau_1\, q\, \bar q$ can pass the 20~GeV cut, leading to more than four jets. Nevertheless, this scenario has a higher chance of reconstructing the two SM bosons. The reason for this is that $\tilde\tau_1$ decay will contribute directly to the signal, while other charged sleptons contribute through a cascade-produced $\tilde\nu_L$. The latter decays into a Higgs 50\% of the time, which has a lower reconstruction efficiency than the $W$ or $Z$. Thus, scenario ST ends up with a slightly higher efficiency overall.

\begin{figure}[tb]
\centering
\includegraphics[width=0.47\textwidth]{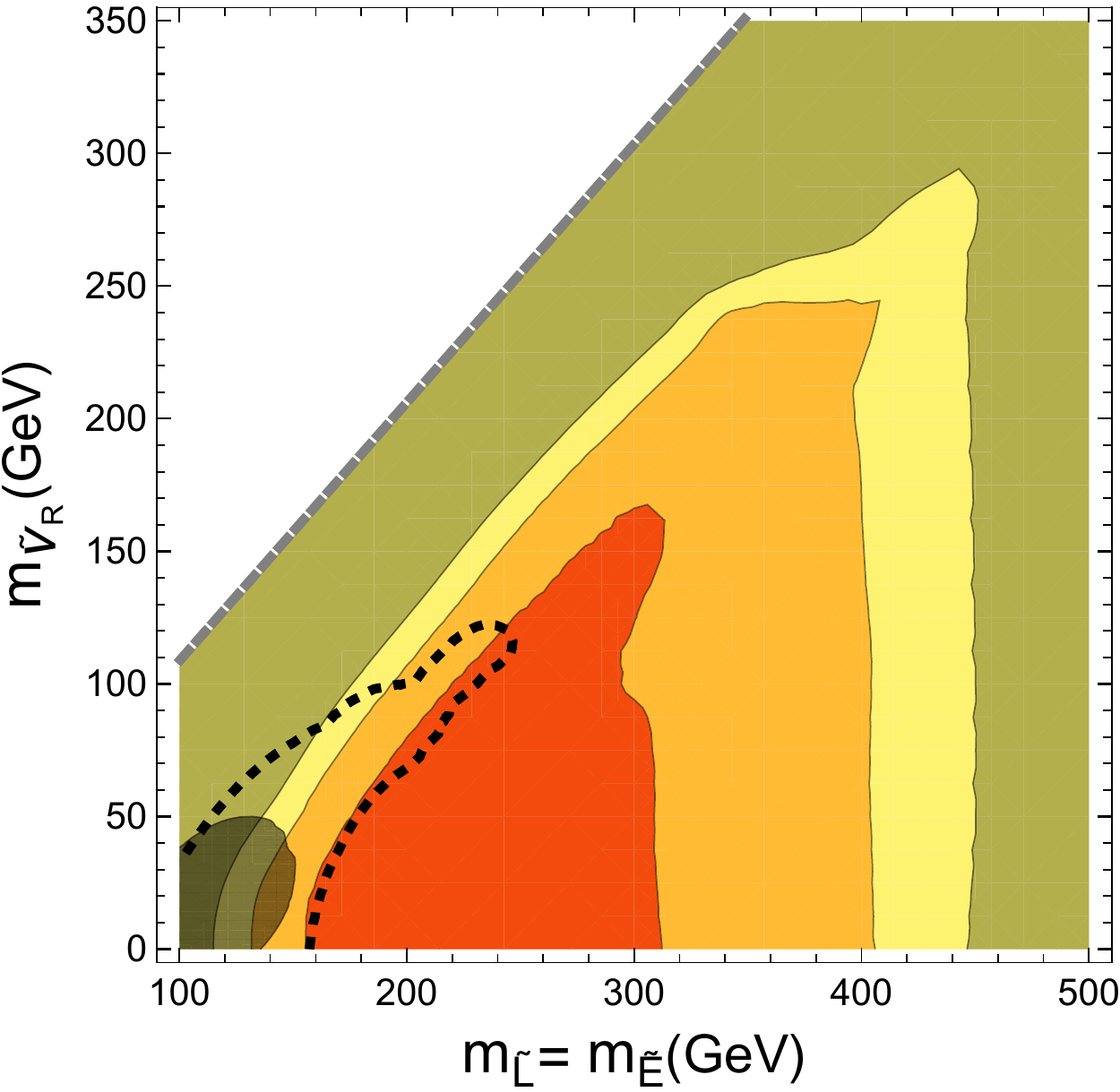} \hfill
\includegraphics[width=0.47\textwidth]{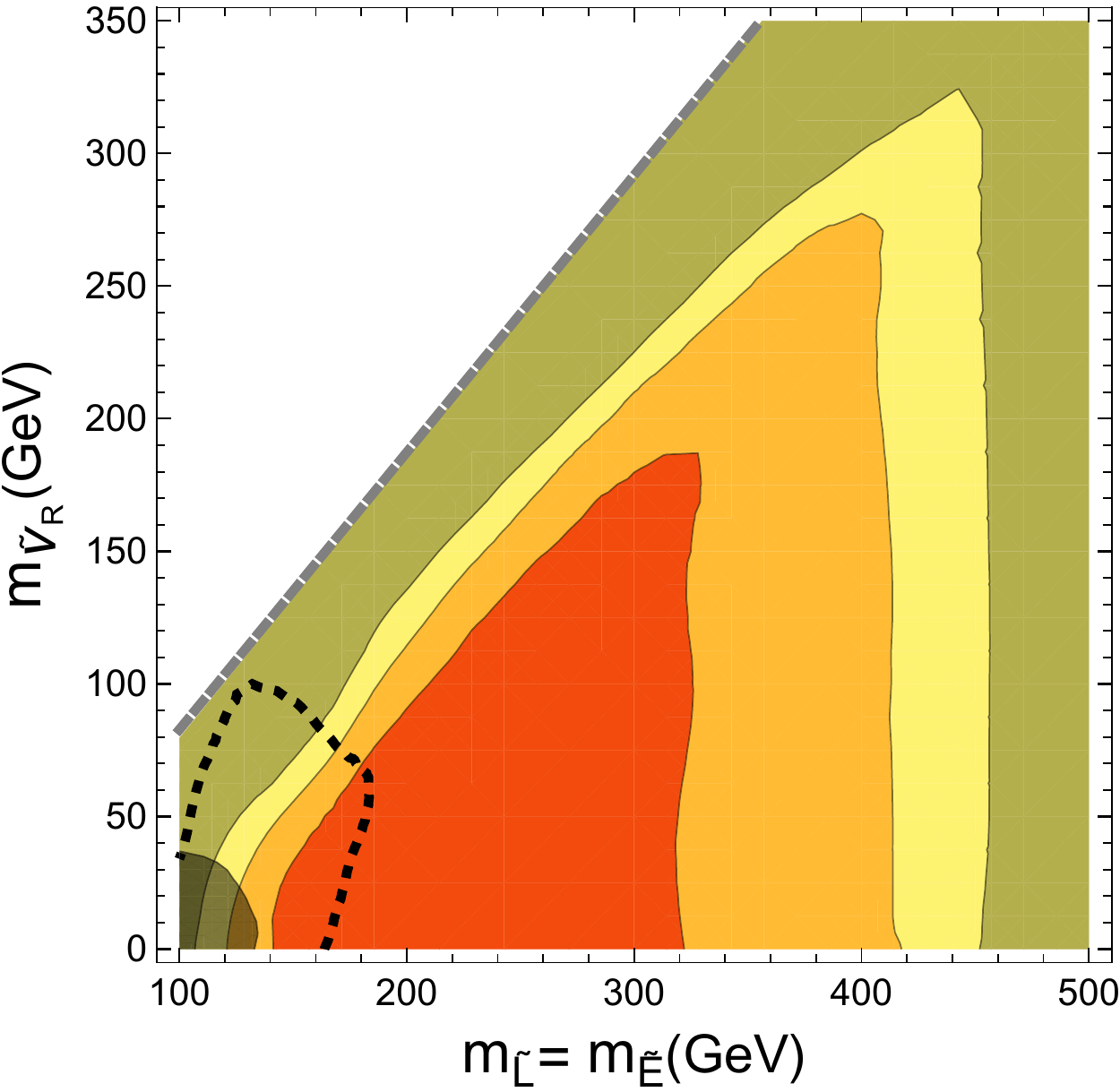} \\ 
\vspace{3mm}
\includegraphics[width=0.47\textwidth]{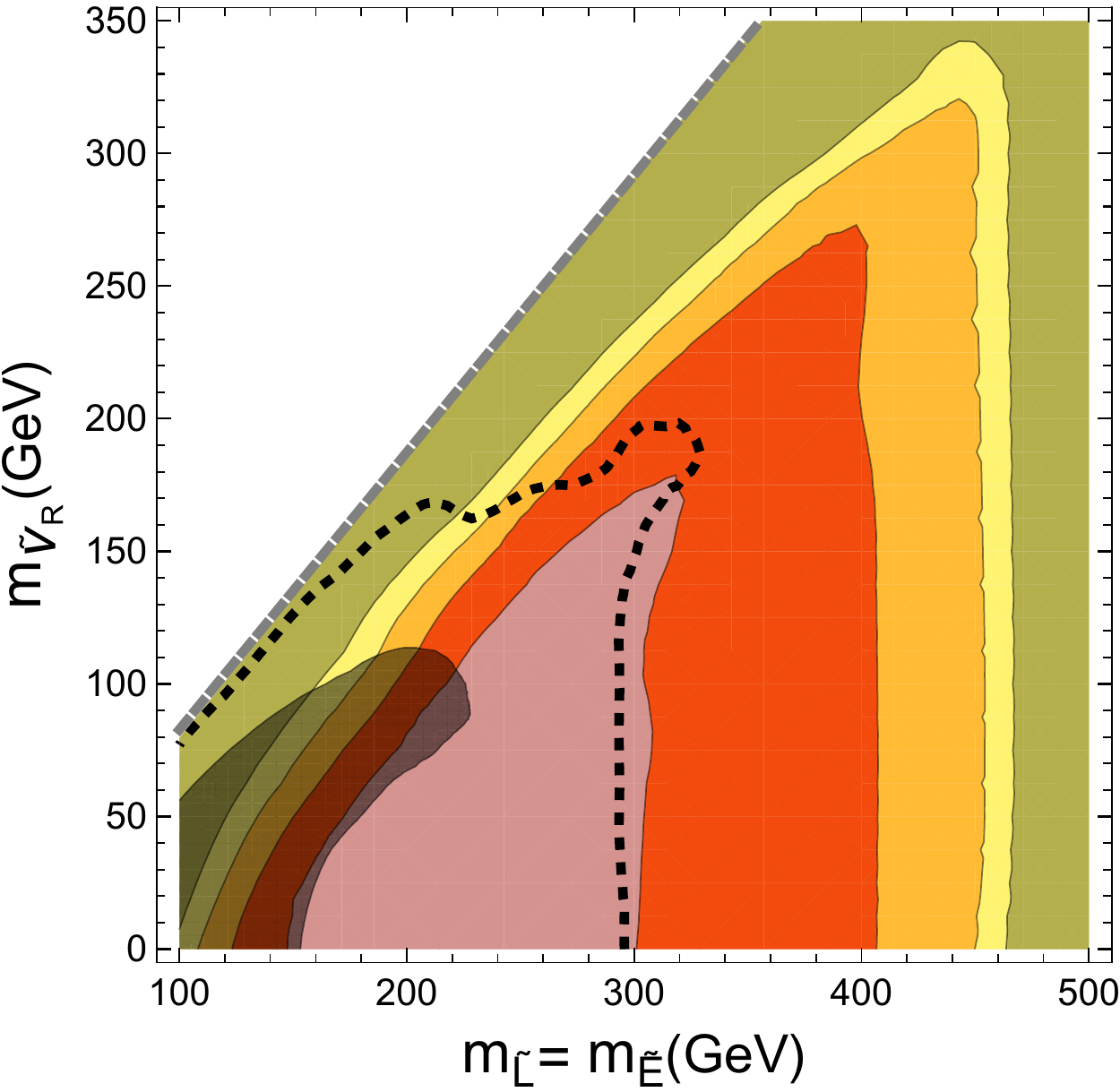} \hfill
\includegraphics[width=0.47\textwidth]{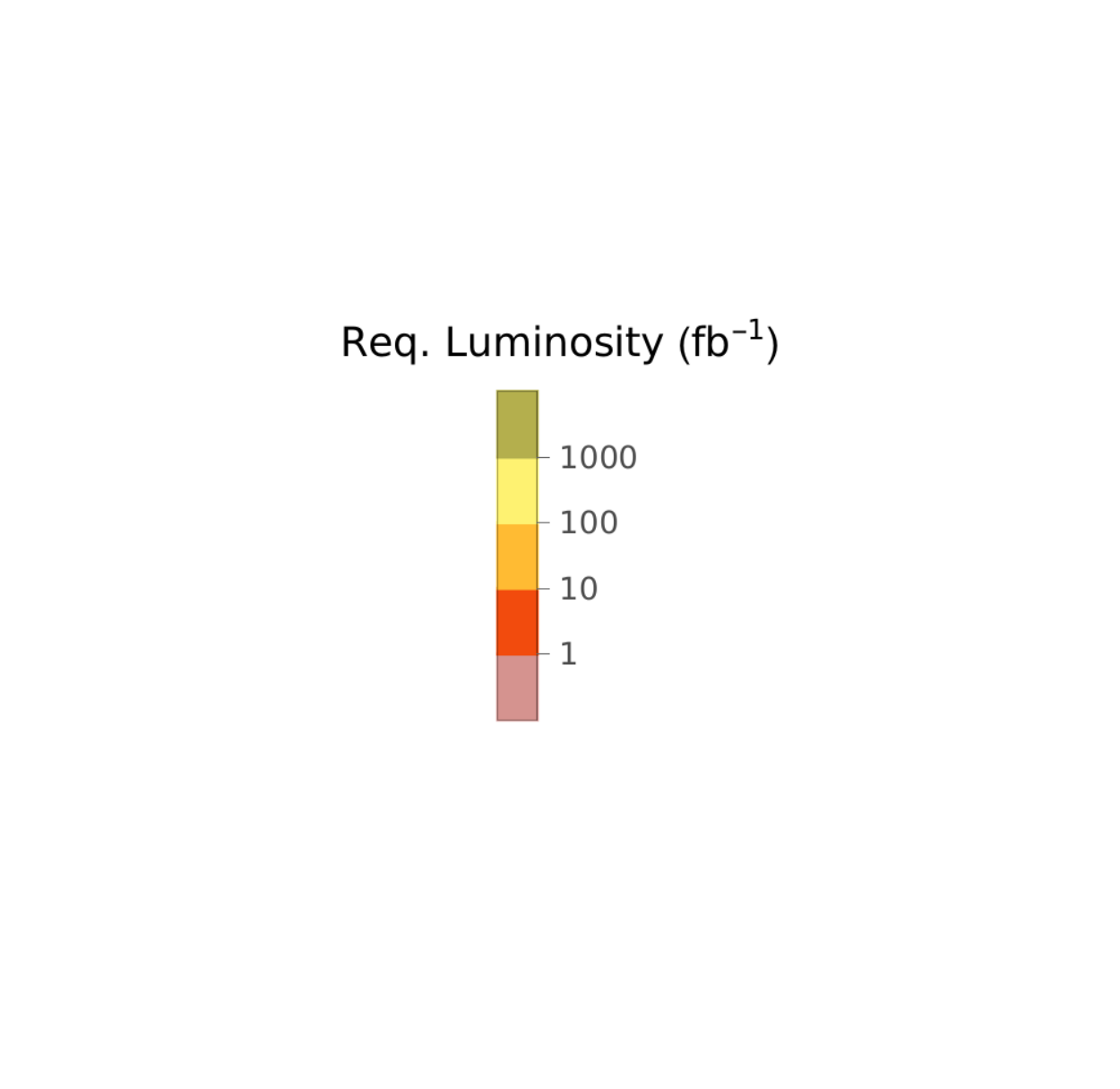}
\caption{Necessary luminosity, in fb$^{-1}$, to achieve $s/\sqrt{b}=5$. Scenarios SE, ST and DEG are shown on the top left, top right, and bottom left panels, respectively. Luminosities lower than 1, 10, 100 and 1000~fb$^{-1}$ are shown in pink, dark orange, light orange and yellow, respectively. The shaded region indicates the current LHC exclusion, interpolated from the data in the previous section. Similarly, the dashed lines indicate the expected LHC reach for 300~fb$^{-1}$.}
\label{fig:all_lum}
\end{figure}
In our benchmark, the efficiency for our signal is about 50 times that of the background, with both signal and background events of the same order. This motivates a scan on $m_{\tilde L}=m_{\tilde E}$ and $m_{\tilde\nu_R}$, which is presented in Fig.~\ref{fig:all_lum}, for scenarios SE, ST and DEG. The figure shows the required luminosity in order to obtain a $5\sigma$ sensitivity. We find that this search can probe a large part of the evaluated parameter space, using a relatively low luminosity. Given the higher efficiency, scenario ST requires slightly less data than scenario SE to achieve $5\sigma$. As expected, for scenario DEG the required luminosity is much smaller, since the total production cross section is larger. Here we find that more than 50\% of the evaluated points would lead to a discovery with less than 100 fb$^{-1}$.

Of course, the required luminosity becomes very large when slepton masses exceed $\sim400$~GeV. This is due to the cross section, which goes to zero when the slepton mass is larger than 500~GeV. We also get a loss in sensitivity when the slepton mass approches $m_{\tilde{\nu}_R}$, as all decay products became soft and no on-shell SM bosons are produced. On the other hand if we have $m_{\tilde{\nu}_R}+100\,{\rm GeV}\lesssim m_{\tilde{L}},m_{\tilde{E}}<400$~GeV, and on-shell SM boson final states, then for a reasonable value of integrated luminosity ($\simeq 100$ fb$^{-1}$) a discovery is ensured.

\section{Mass Reconstruction}
\label{sec:mass_reconstruction}

\subsection{Endpoint Method}
\label{sec:method}

Assuming a slepton discovery is made at the ILC, it is desirable to extract as much information as possible regarding the new particles. In this section we will evaluate the endpoint method for mass reconstruction presented in~\cite{Suehara:2009bj} (see also~\cite{Kafer:2009gm,Li:2010mq,Alster:2011he} and Ch.~11 in~\cite{Chera:2018hmr}), for our model. This method was designed to reconstruct chargino and neutralino masses on a specific simplified model~\cite{Battaglia:2006bv}, with a $\tilde\chi^0_2-\tilde\chi^\pm$ pair decaying into $Z\,\tilde\chi^0_1$ and $W\,\tilde\chi^0_1$ states. Thus, we consider it is particularly adequate for scenario ST, with both sleptons decaying into on-shell SM bosons ($W,\,Z,\,h$) and a $\tilde\nu_R$, with branching ratios as in Fig.~\ref{fig:BR}. We note for completeness that an endpoint method had also been used in \cite{Dima:2001jr,Battaglia:2013bha} in case of the standard slepton decays into leptons and electroweakinos.

An alternative procedure for mass reconstruction, outside of the scope of this work, is the threshold analysis~\cite{Mizukoshi:2001nc,Feng:2001ce,Martyn:2003av,Blochinger:2002zw,MoortgatPick:2005cw}. Here, the production cross section is measured as $\sqrt s$ is varied, with the shape giving information regarding the mass and spin of the produced particles.

The endpoint method aims to reconstruct SUSY masses using the maximum and minimum measured values of the SM boson energy spectrum. The latter values are called the \textit{endpoints} of the distribution. In order to reconstruct $m_{\tilde\ell}$, $m_{\tilde\nu_{L}}$ and $m_{\tilde\nu_R}$, we require the energy of the outgoing SM boson on the rest frame of the decaying slepton. From 2-body kinematics, it is given by:
\begin{equation}
\label{eq:ebprime}
    E'_B=\dfrac{m_{\tilde{\ell}}^2+m_B^2-m_{\tilde\nu_R}^2}{2 m_{\tilde{\ell}}}\quad,
\end{equation}
where $m_{\tilde \ell}$ and $m_B$ denote the masses of the relevant slepton and corresponding SM boson, respectively. Even though included in our simulation, in this analysis we neglect possible energy losses due to beamstrahlung and ISR, and consider $E_{\tilde{\ell}}=E_{\rm beam}=500$~GeV. 
The lower and upper endpoints are then given by boosting into the lab frame:
\begin{eqnarray}
\label{eqn:endpoints1}
    E_{B-}&=&E'_B\frac{E_{\rm beam}}{m_{\tilde{\ell}}} -\sqrt{E_B^{\prime 2} -m_B^2}\frac{\sqrt{E_{\rm beam}^2-m_{\tilde{\ell}}^2}}{m_{\tilde{\ell}}} \\
    \label{eqn:endpoints2}
    E_{B+}&=&E'_B\frac{E_{\rm beam}}{m_{\tilde{\ell}}} +\sqrt{E_B^{ \prime 2} -m_B^2}\frac{\sqrt{E_{\rm beam}^2-m_{\tilde{\ell}}^2}}{m_{\tilde{\ell}}}
\end{eqnarray}
From the equations above one can solve for $m_{\tilde\ell}$ and $E'_B$:
\begin{eqnarray}
E'_B &=& \frac{1}{\sqrt2}\sqrt{(E_{B+}\, E_{B-} + m_B^2)\pm \sqrt{(E_{B+}^2-m_B^2)(E_{B-}^2-m_B^2)}} \\
m_{\tilde{\ell}} &=& \frac{2E_{\rm beam}}{E_{B+}+E_{B-}}E'_B~,
\end{eqnarray}
which connects the endpoints with the slepton masses. However, there are two values of $E'_B$ that are consistent with the measurement, leading to a degeneracy in the mass determination. In order to solve the degeneracy, we need to use at least two datasets, that is, two sets of data involving different SM bosons, thus leading to different endpoints. The point is that for each dataset, this procedure allows us to also deduce the R-sneutrino mass:
\begin{equation}
    m_{\tilde\nu_R}=\sqrt{m_{\tilde{\ell}}^2+m_B^2-2E'_B m_{\tilde{\ell}}}
\end{equation}
Thus, the correct sign for $E'_B$ is determined by requiring that the reconstructed $m_{\tilde\nu_R}$ is equal for all slepton decays.

If the $\tilde\nu_R$ mass is known, one can avoid using one of the endpoints to obtain the slepton mass, again having two solutions. For example, using only the upper endpoint, we find:
\begin{equation}
\label{eq:massupperendp}
m_{\tilde\ell}^2=2(E_{\rm beam}-E_{B+}) E_{B+} + m_B^2 + m^2_{\tilde\nu_R}\pm
2\sqrt{(E_{B+}^2 - m_B^2) ((E_{\rm beam} - E_{B+})^2 -m^2_{\tilde\nu_R})}
\end{equation}
This equation can be relevant in scenarios in which the lower endpoint is very close to the SM boson mass. In this case, the width of the boson could make it difficult to resolve $E_{B-}$ experimentally~\cite{Suehara:2009bj}. Another reason for using only the upper endpoint comes when the SUSY background modifies the lower endpoint position. Here, the correct sign for $m_{\tilde\ell}^2$ can be fixed by choosing the theoretical value of $E_{B-}$, reconstructed from $E_{B+}$ and $m_{\tilde\nu_R}$, closest to the boson mass.

In order to obtain the endpoints, $E_{B-}$ and $E_{B+}$, we follow the recipe described in~\cite{Suehara:2009bj}. First, we group all events into three datasets ($W$-like, $Z$-like and $h$-like), based on the type of final state (light jets or b-jets) and the reconstructed invariant mass ($m_W,\,m_Z,\,m_h$). As in Sec.~\ref{sec:sensitivity}, we only consider decays of SM bosons into a dijet. For $W$ boson pairs, we form the dijets by requiring exactly four light jets, with the invariant mass of each dijet reconstructing $m_W$. For $h$ boson pairs, we follow an analogous procedure, but use b-jets instead. For $Z$ boson pairs, we use both light jets and b-jets. Notice that this selection is stricter than the one in Sec.~\ref{sec:sensitivity}, since we require both bosons to have the same mass.

In more detail, for the case with four light jets, we consider all possible jet combinations, using the following discriminating variables:
\begin{equation}
\chi^2_W(m_1,m_2)=\dfrac{(m_1-m_W)^2+(m_2-m_W)^2}{\sigma^2}
\end{equation}
\begin{equation*}
\chi^2_Z(m_1,m_2)=\dfrac{(m_1-m_Z)^2+(m_2-m_Z)^2}{\sigma^2}
\end{equation*}
with $m_1$ and $m_2$ the dijet masses and $\sigma=5$\,GeV. An event is included in the $W$-like dataset ($\tilde{\tau}_1$ decay) if $\chi_W^2<4$ and $\chi_Z^2>5$, or in the $Z$-like dataset ($\tilde{\nu}_{L\tau}$ decay) if $\chi_W^2>4$ and $\chi_Z^2<2$. Similarly, for four b-jets we define:
\begin{equation}
\chi^2_h(m_1,m_2)=\dfrac{(m_1-m_h)^2+(m_2-m_h)^2}{\sigma^2}
\end{equation}
\begin{equation*}
\chi^2_Z(m_1,m_2)=\dfrac{(m_1-m_Z)^2+(m_2-m_Z)^2}{\sigma^2}
\end{equation*}
An event is included in the $h$-like dataset ($\tilde{\nu}_{L\tau}$ decay) if $\chi_h^2<4$ and $\chi^2_Z>5$, or in the $Z$-like dataset ($\tilde{\nu}_{L\tau}$ decay) if $\chi_h^2>4$ and $\chi_Z^2<4$. In this case we relax the $\chi^2_Z$ requirement in comparison to that for light jets due to the larger mass difference between the $Z$ and $h$ bosons.

Having separated the events into the three datasets, for each group we fit the reconstructed energy spectrum, following the steps described in~\cite{Suehara:2009bj}:
\begin{enumerate}
    \item We take the MC events corresponding to the SM background, and use them to fit the six parameters of the following distribution:
    \begin{equation}
    \label{eq:SMback}
        f_{SM}(E;\,E_{\rm SM-},\,a_{0-2},\,\sigma_{\rm SM},\,\Gamma_{\rm SM})=\int_{E_{\rm SM-}}^\infty (a_2 E^{\prime2}+a_1 E'+a_0)\,V(E'-E,\sigma_{\rm SM},\Gamma_{\rm SM})\,dE'\quad,
    \end{equation}
    with $E$ the boson energy, and $V(E'-E,\sigma_{\rm SM},\Gamma_{\rm SM})$ a Voigt function of resolution $\sigma_{\rm SM}$ and width $\Gamma_{\rm SM}$. The second order polynomial determines the shape of the distribution, and $E_{\rm SM-}$ adjusts the threshold position.
    
    \item Using the fitted parameters, we generate one hundred new datasets of SM background following the $f_{SM}$ distribution. Statistical errors were implemented by modifying the number of events on each bin by a random number following a Poissonian distribution around the center of the bin.
    
    \item For each SM dataset, we fit the sum of the SUSY and SM spectra into a new distribution:
      \begin{eqnarray}
        f(E;\,E_{B-},\,E_{B+},\,b_{0-2},\sigma_1,\Gamma_1)&=&f_{SM}(E;\,E_{\rm SM-},a_{0-2},\sigma_{\rm SM},\Gamma_{\rm SM}) \nonumber \\
        &&+\int_{E_{B-}}^{E_{B+}} (b_2 E^{\prime2}+b_1 E'+b_0)V(E'-E,\sigma_1,\Gamma_1)dE' \nonumber \\
    \end{eqnarray}
    In~\cite{Suehara:2009bj}, the parameter $\sigma_1$ is allowed to have linear dependence on $E$, but we opt to use constant $\sigma_1$ as it gives us a better fit.
\end{enumerate}

The upper and lower endpoint are then given by the averaged values of $E_{B-}$ and $E_{B+}$, respectively, with errors given by the standard deviation. In the case of b-jets the SM background is negligible, so we divide the data in subsets and apply the fit to each sample \cite{Berggren:2015qua}. We then take the average and standard deviation to estimate their error. Finally, we introduce the endpoints in equations~(\ref{eq:ebprime})-(\ref{eq:massupperendp}) to determine the masses, using error propagation to analytically estimate the uncertainties. The expected spectra for the boson energies are boxlike, with deviations attributed to (1) the massive boson finite widths, (2) detector resolution, including the reconstruction efficiency via jets, and (3) the presence of ISR and beamstrahlung~\cite{Alster:2011he}. In models in which the SUSY particles decay to electrons or muons the edges are considerably more steep and the distribution is markedly boxlike \cite{Tsukamoto:1993gt,Abe:2001wn}. 

\subsection{Results}
\label{sec:results}

In the following we evaluate the method at $\sqrt s=1$~TeV for our three scenarios, fixing the light slepton soft masses to 300~GeV and the soft R-sneutrino mass to 100~GeV. Once these are given, the expected endpoints are set to specific values, reported on the last column of Table~\ref{table:endpoints}.

For all scenarios, we generate 500~fb$^{-1}$ of data with type \textbf{B} polarization. In addition, due to the small number of $h$-like events, we add an additional 500~fb$^{-1}$ to this channel only, this time with type \textbf{L} polarization. The latter polarization is taken since the number of $h$-like events is enhanced by a factor 1.48 without increasing backgrounds. In the case of b-jets the SM background is negligible.

\begin{table}[t]
\centering
\setlength{\tabcolsep}{1.5em}
{\begin{tabular}{| c | c | c|  c|c|} 
\hline
Endpoint & SE & ST & DEG & Theory\\
\hline\hline
$E_{W-}$ (GeV) & $95.24\pm3.77$ & $80.49\pm0.43$ & $81.52\pm0.64$ & 80.88 / 80.41 \\
\hline
$E_{W+}$ (GeV) & $347.38\pm18.35$ & $398.11\pm1.11$ & $398.59\pm1.15$ & 399.81 / 399.09 \\
\hline
$E_{Z-}$ (GeV) & $91.90\pm 0.40$ & $92.66\pm0.65$ & $92.38\pm0.84$ & 91.66 \\
\hline
$E_{Z+}$ (GeV) & $397.52\pm1.79$ & $397.85\pm1.82$ & $397.92\pm1.75$& 398.53 \\
\hline
$E_{h-}$ (GeV) & $136.89\pm1.45$ &$137.05\pm 1.69$ & $137.05\pm1.01$ & 137.25 \\
\hline
$E_{h+}$ (GeV) & $396.09\pm1.18$ &$396.00\pm 1.29$ & $395.70\pm0.61$ & 395.65 \\
\hline
\end{tabular}}
\caption{The different endpoints reconstructed for each scenario. The last column shows our theoretical expectation. For $E_{W-}$ and $E_{W+}$ we report two theoretical endpoints, the first one corresponding to the SE scenario, and the second to the ST and DEG scenarios.}
\label{table:endpoints}
\end{table}

\begin{figure}[t]
	\centering
\includegraphics[width=0.49\textwidth]{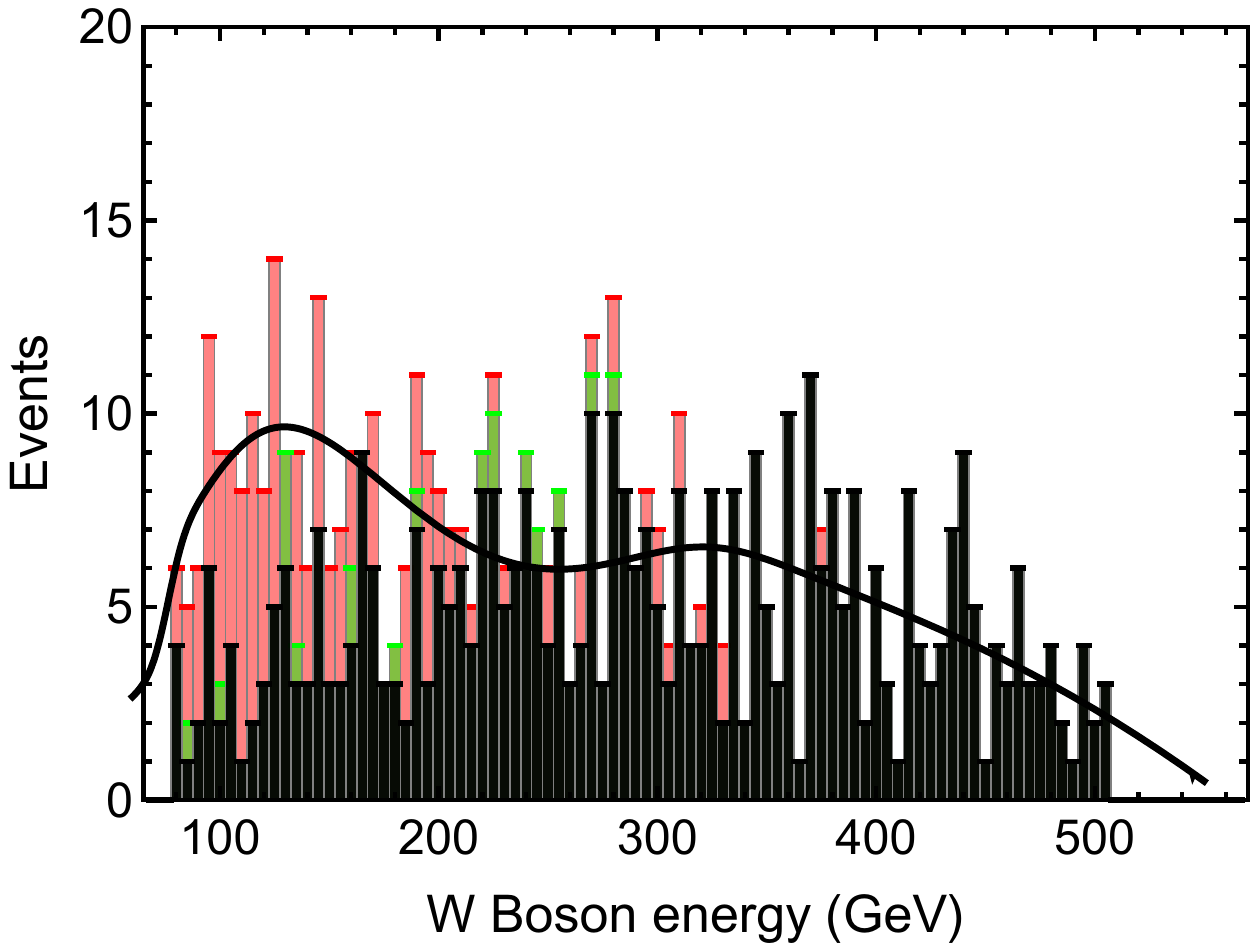} \hfill
\includegraphics[width=0.49\textwidth]{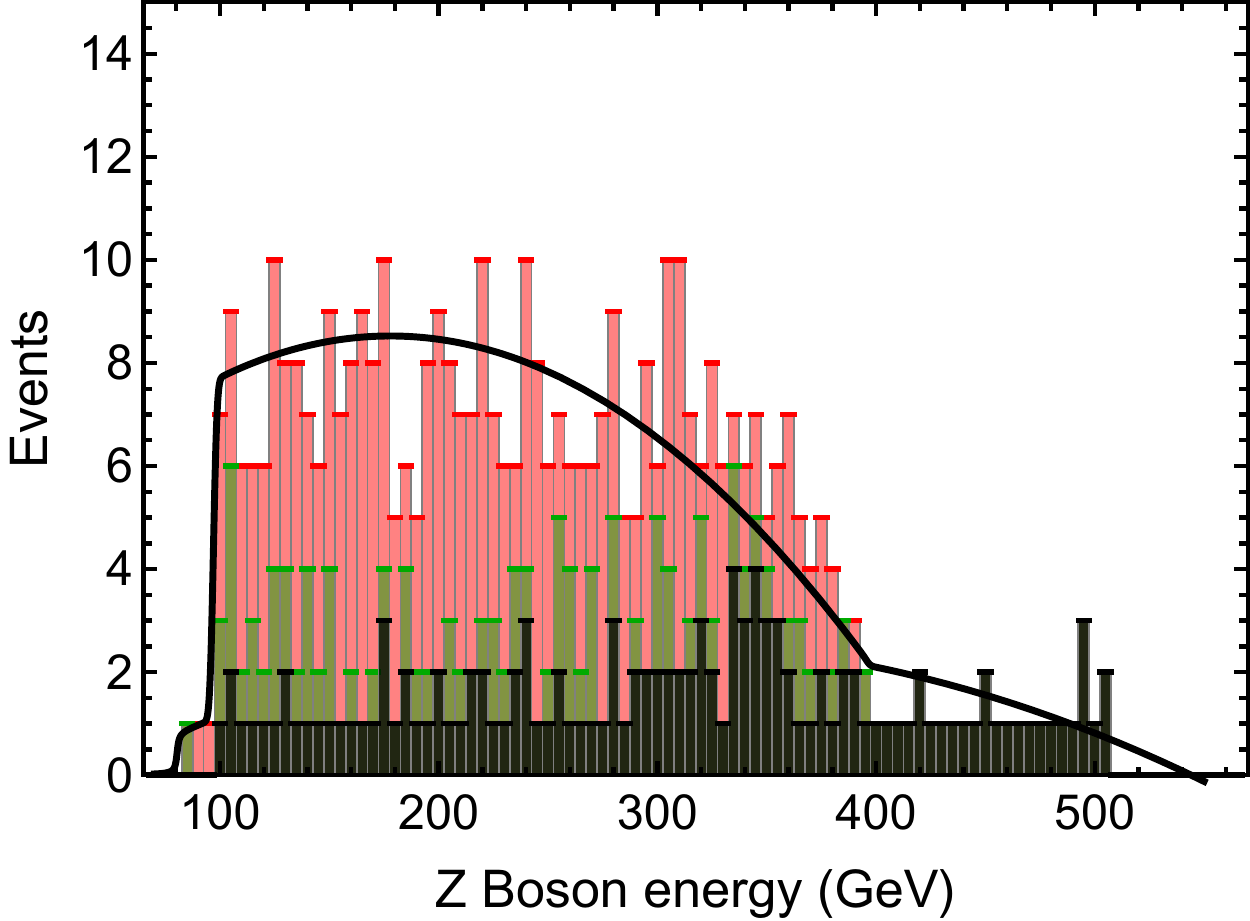} \\ \vspace{3mm}
\includegraphics[width=0.49\textwidth]{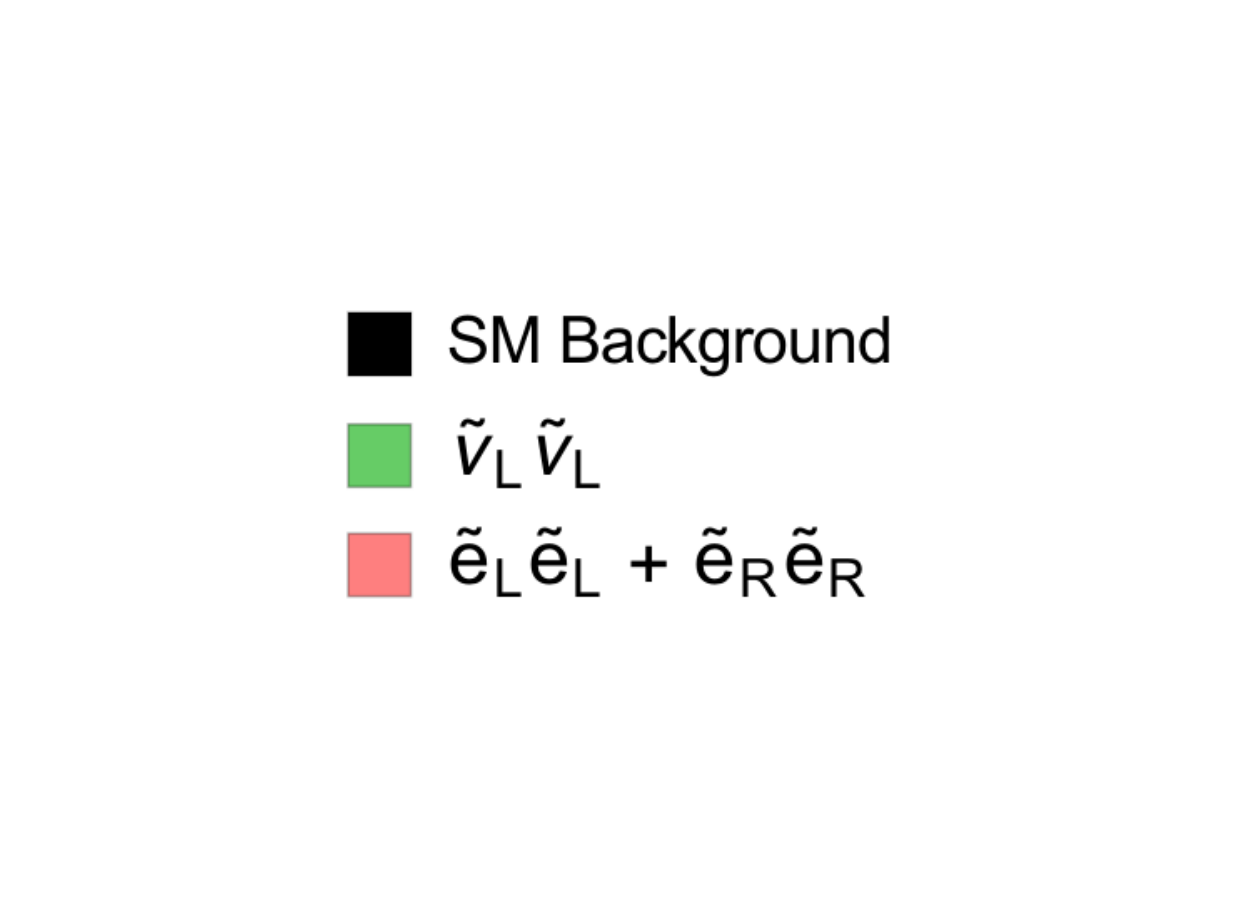} \hfill
\includegraphics[width=0.49\textwidth]{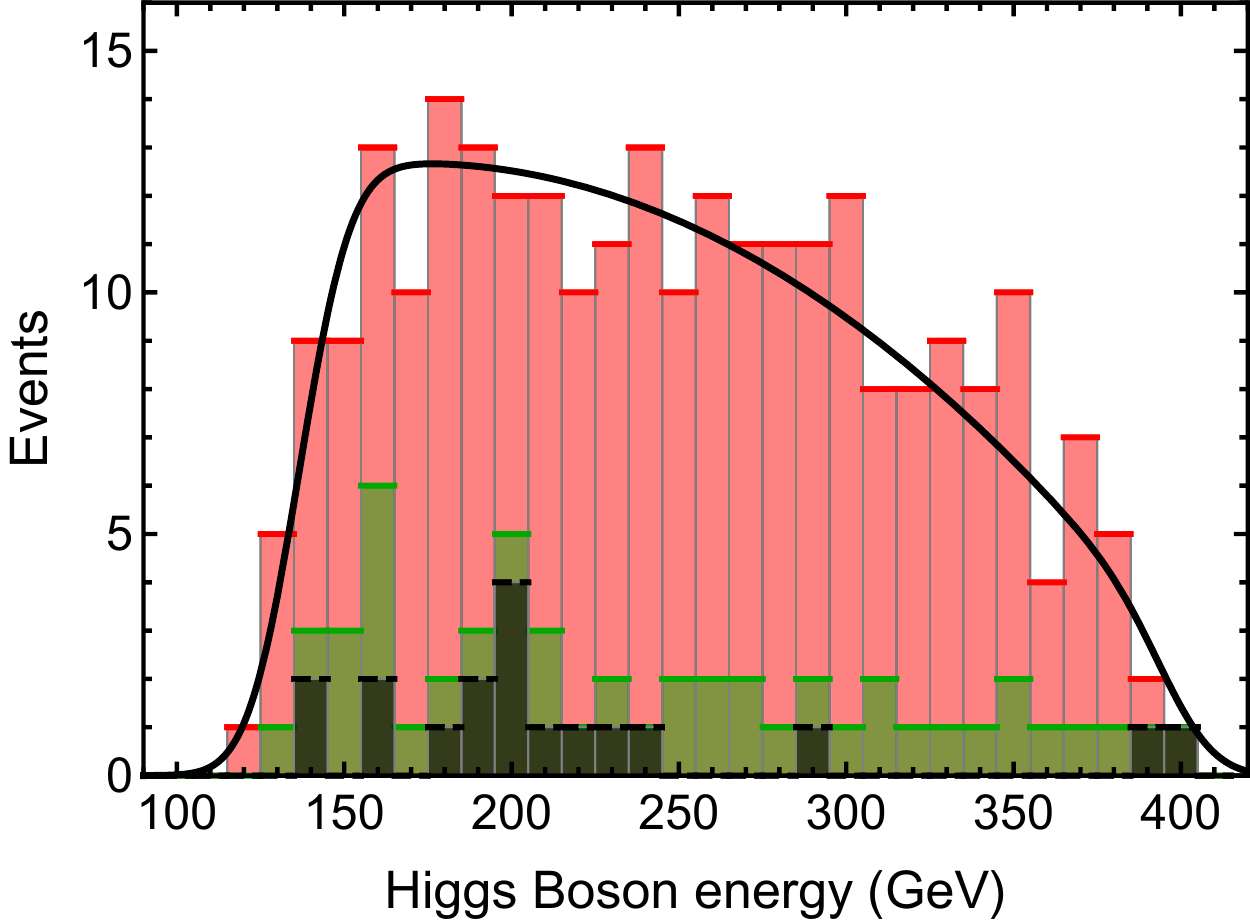} 
\caption{Energy spectra of reconstructed $W$ (top left), $Z$ (top right), and $h$ bosons (bottom right) for Scenario SE. SM background is shown in black, $\tilde\nu_L$ contributions in green, $\tilde e_{L,R}$ cascade events in red. The solid line is the result of the fit.}
\label{fig:energyA}
\end{figure}
The fits for $W$-like, $Z$-like and $h$-like events in Scenario SE are shown in Fig.~\ref{fig:energyA}, with the reconstructed endpoints reported on the first row of Table~\ref{table:endpoints}. Since the $\tilde e_{L,R}$ cascade into $\tilde\nu_L$, we do not obtain any useful information from the $W^\pm$ endpoints. All of the $W$-like events come either from SM background, or incorrectly identified $Z$ bosons coming either from directly produced $\tilde\nu_L$ or from the $\tilde e_{L,R}$ cascade. Thus, due to the low signal statistics, it is not possible to reconstruct the correct endpoints with this data. This is reflected on Table~\ref{table:endpoints}.

Interestingly, the $Z$-like and $h$-like datasets receive an overwhelming contribution from the $\tilde e_{L,R}$ cascade ($\simeq 80\%$ of the total SUSY events). This is consistent with the different cross-sections, as shown in Fig.~\ref{fig:cross}. In spite of this, one can still reconstruct the correct endpoints. In fact, one can consider $e^+e^-\to\tilde e^+\tilde e^-$ as an alternative production channel for $\tilde\nu_L$. The endpoint method works as long as the L-sneutrino energy can still be taken equal to $E_{\rm beam}$ and, given the requirement of having only four high-energy jets, we find that this can be taken to a good approximation. Thus, we consider both $Z$-like and $h$-like datasets to be reliable, and proceed with the reconstruction of $m_{\tilde{\nu}_L}$ and $m_{\tilde{\nu}_R}$ masses using the standard method. Results are shown on the SE column in Table~\ref{table:results}, where we see that the best-fit values for the masses lie within less than $1\%$ of the true ones. On the other hand, as expected, it is not possible to correctly reconstruct $m_{\tilde{\ell}_1}$ at all.

\begin{figure}[t]
	\centering
\includegraphics[width=0.49\textwidth]{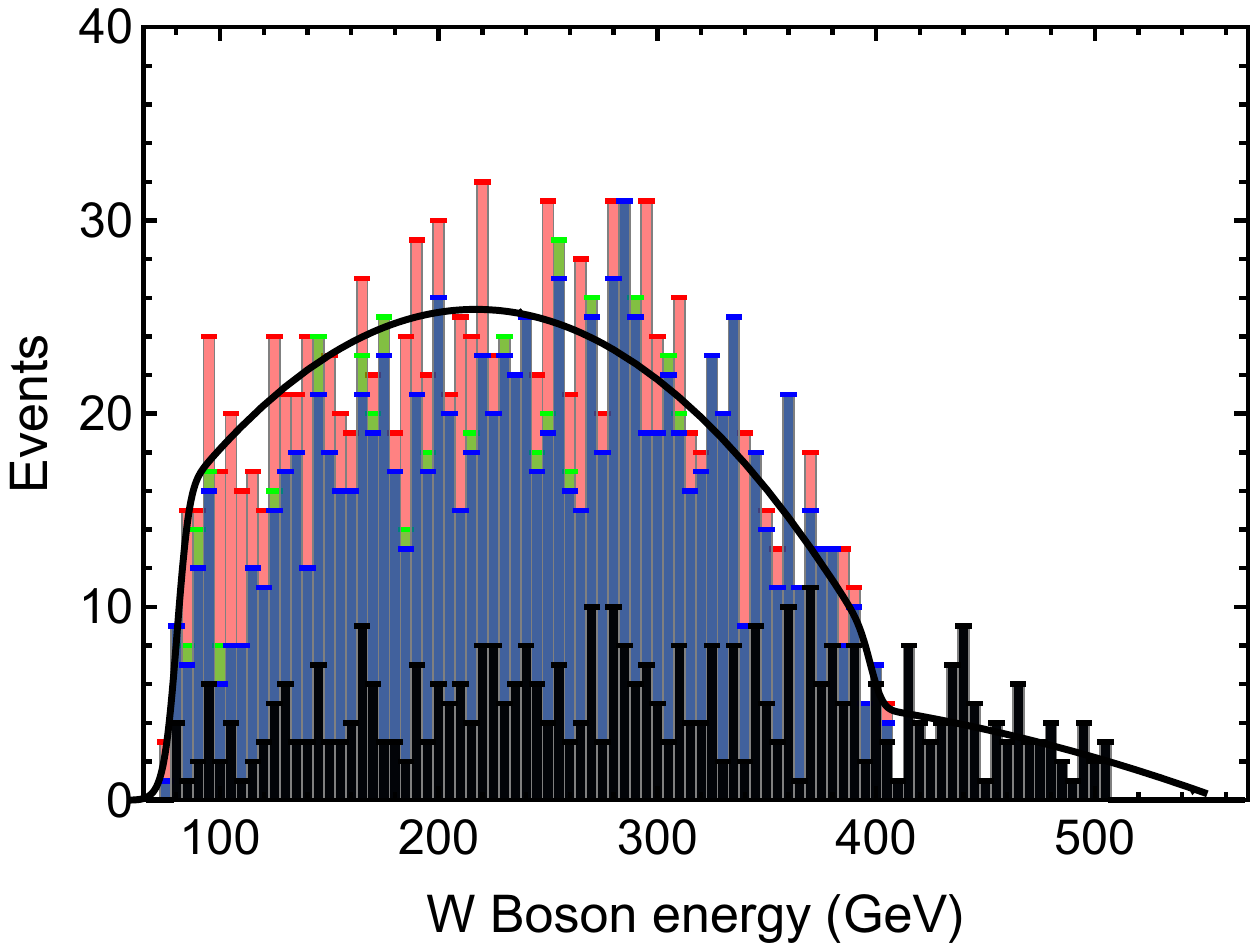} \hfill
\includegraphics[width=0.49\textwidth]{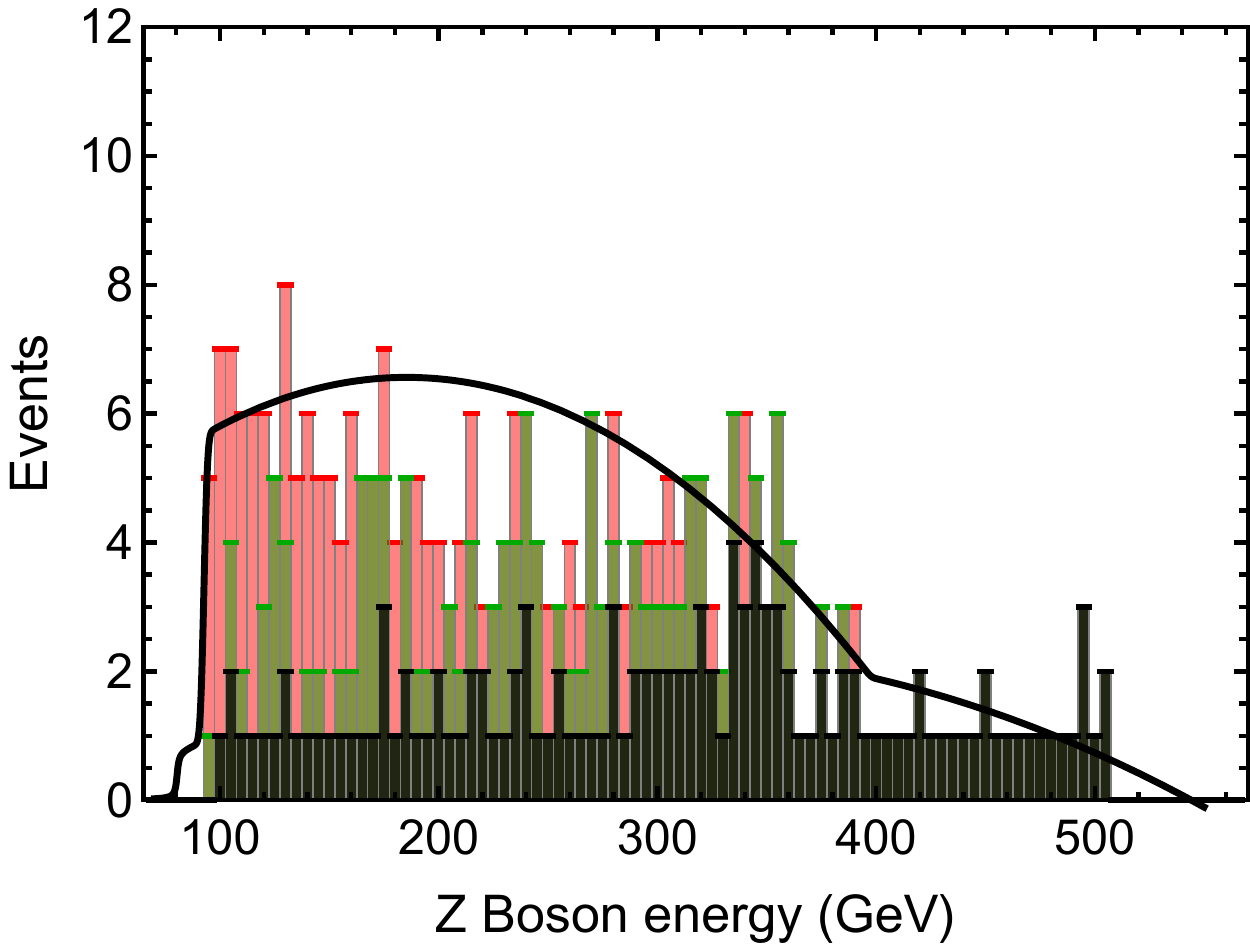}\\\vspace{3mm}
\includegraphics[width=0.49\textwidth]{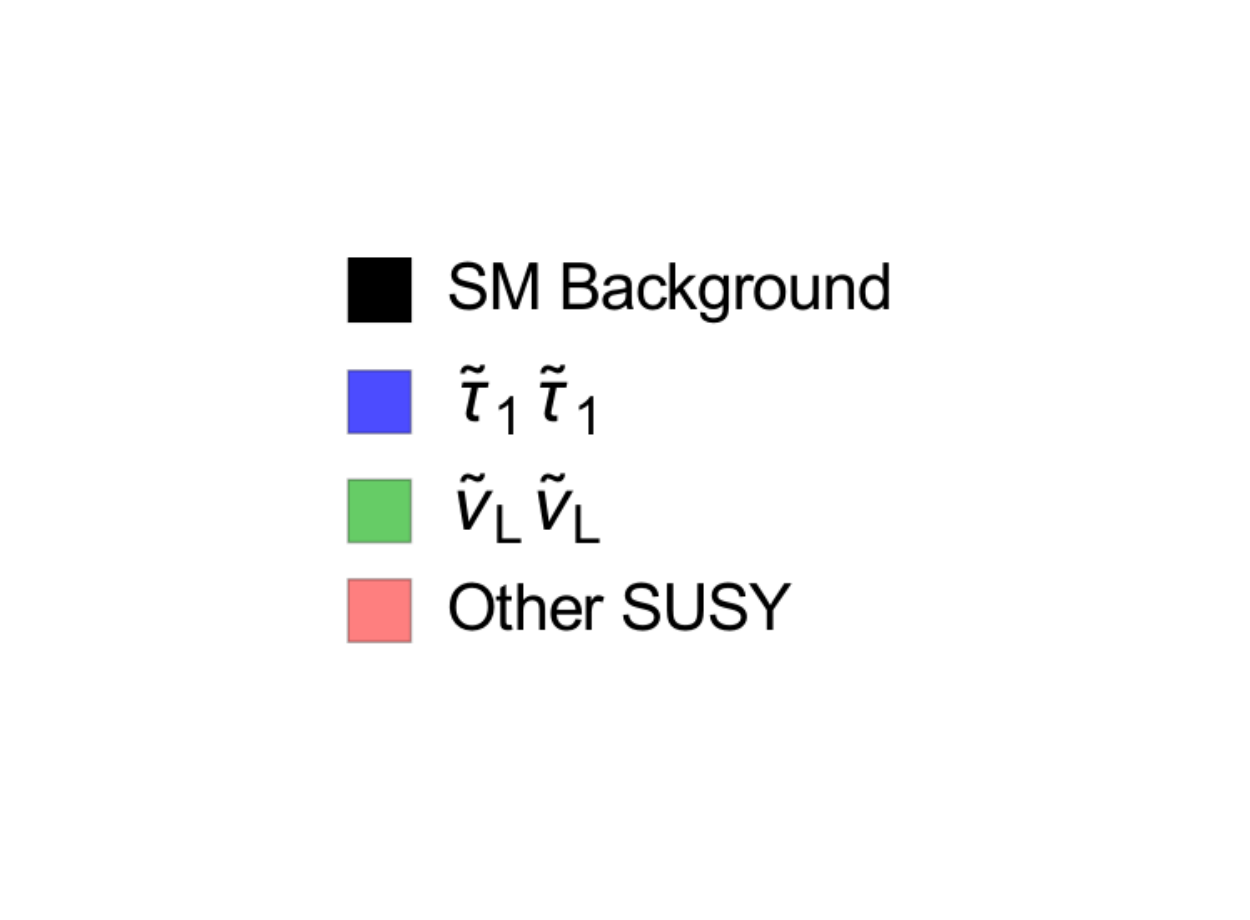} \hfill
\includegraphics[width=0.49\textwidth]{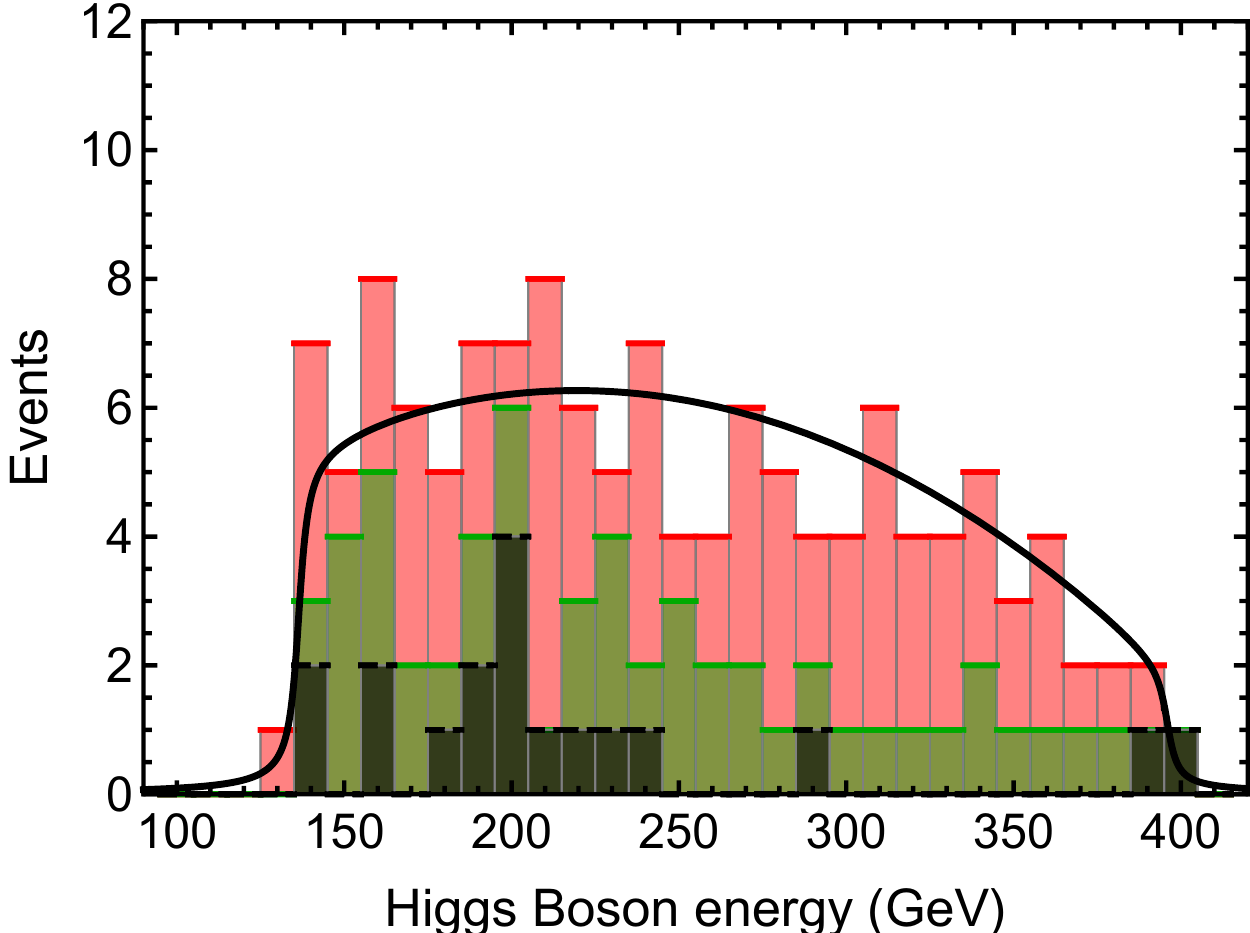}
\caption{Energy spectra of reconstructed $W$ (top left), $Z$ (top right), and $h$ bosons (bottom right) for Scenario ST. SM background is shown in black, and $\tilde\nu_L$, $\tilde\tau_1$, and $\tilde\tau_2$ contributions in green, blue, and red, respectively. The solid line is the result of the fit.}
\label{fig:energyB}
\end{figure}
For Scenario ST we get signal events in all $W$-like, $Z$-like and $h$-like datasets. The boson energy spectrum, along with the fit, can be seen in Fig.~\ref{fig:energyB}. This time the $\tilde\tau_1$ contribution to $W$-like events is dominant, exceeding the SM background by a factor 1.9. Here we find again a small contamination from misidentified $\tilde\nu_L$ and $\tilde\tau_2$ decays. Similar to the SE scenario, the $Z$-like and $h$-like datasets have a large contribution from $\tilde\tau_2$ cascade decays.

In order to reconstruct the masses, we obtain both upper and lower endpoints of the $W$-like and $h$-like datasets, following then the standard method. We find that this strategy works slightly better than using $W$-like and $Z$-like, or $Z$-like and $h$-like datasets, mainly due to the larger number of events. An analysis using $W$-like and $Z$-like events would give similar results. The reconstructed endpoints and masses are shown on the ST column of Tables~\ref{table:endpoints} and~\ref{table:results}, respectively. We see that the best-fit values for $m_{\tilde\tau_1}$ and $m_{\tilde\nu_R}$ are within $\sim1\%$ of the theoretical value, although here the precision for $m_{\tilde\nu_L}$ is slightly lower than for the SE scenario. This is attributed to the lower number of $h$-like events.

\begin{figure}[t]
	\centering
\includegraphics[width=0.49\textwidth]{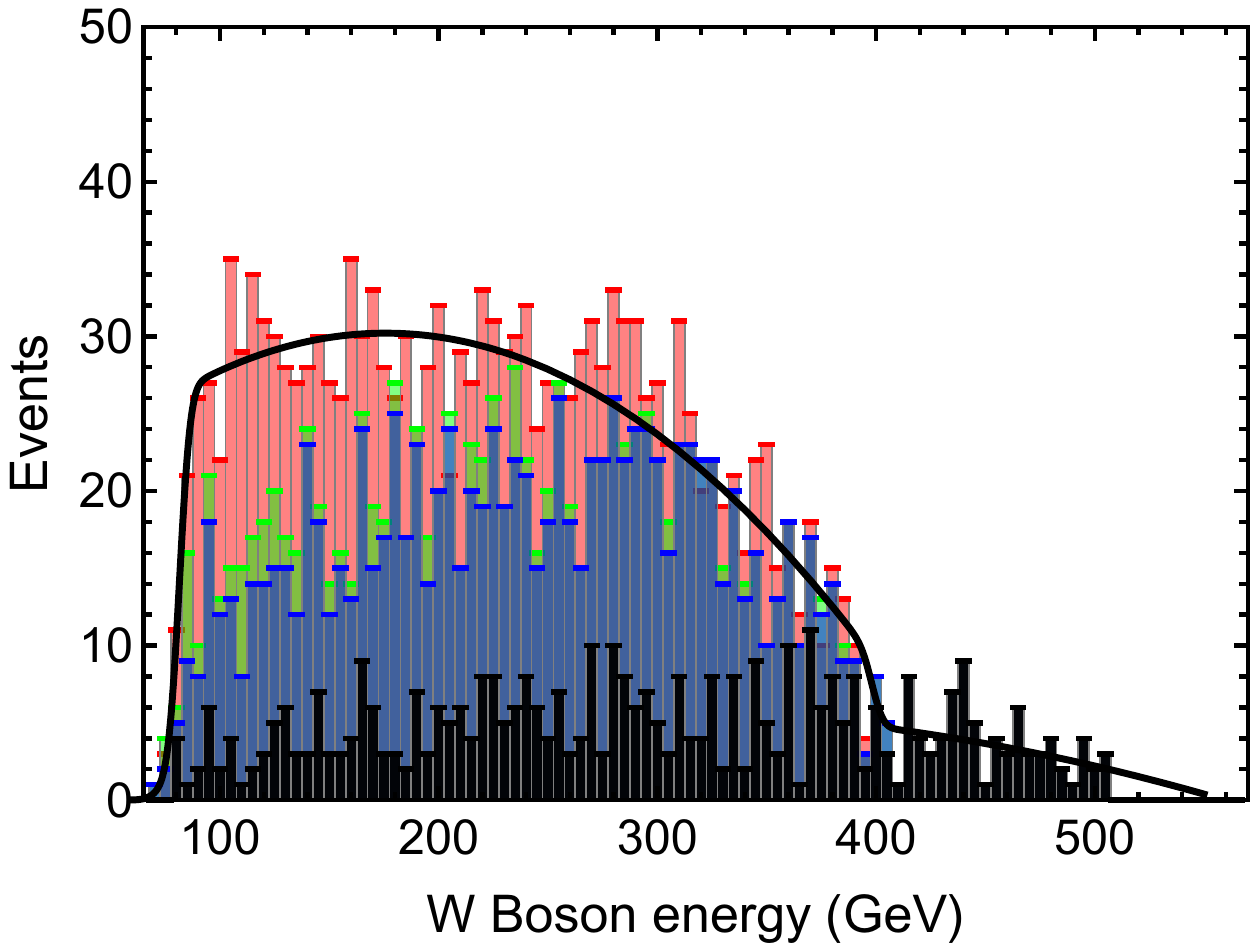} \hfill
\includegraphics[width=0.49\textwidth]{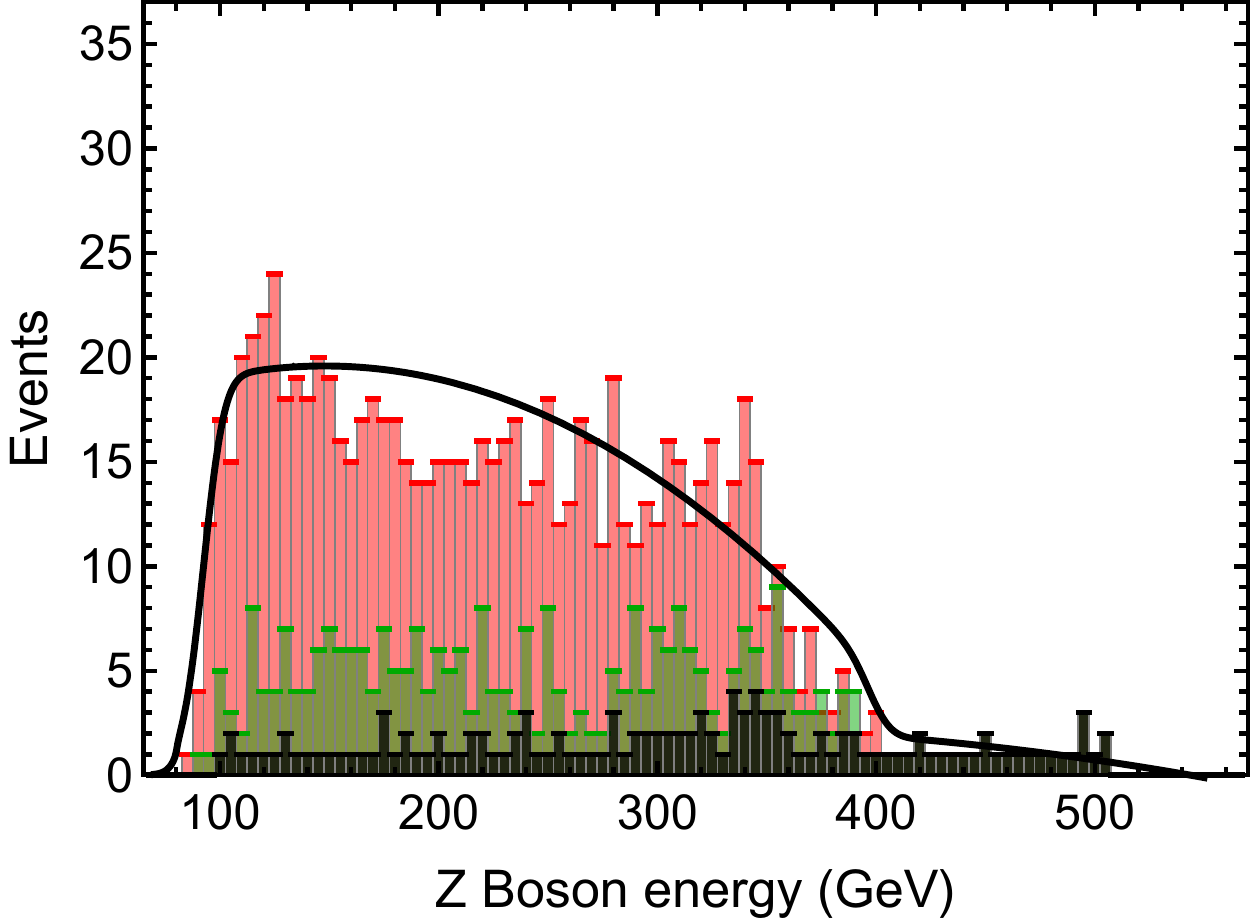} \\ \vspace{3mm}
\includegraphics[width=0.49\textwidth]{brs_mass_st.pdf} \hfill
\includegraphics[width=0.49\textwidth]{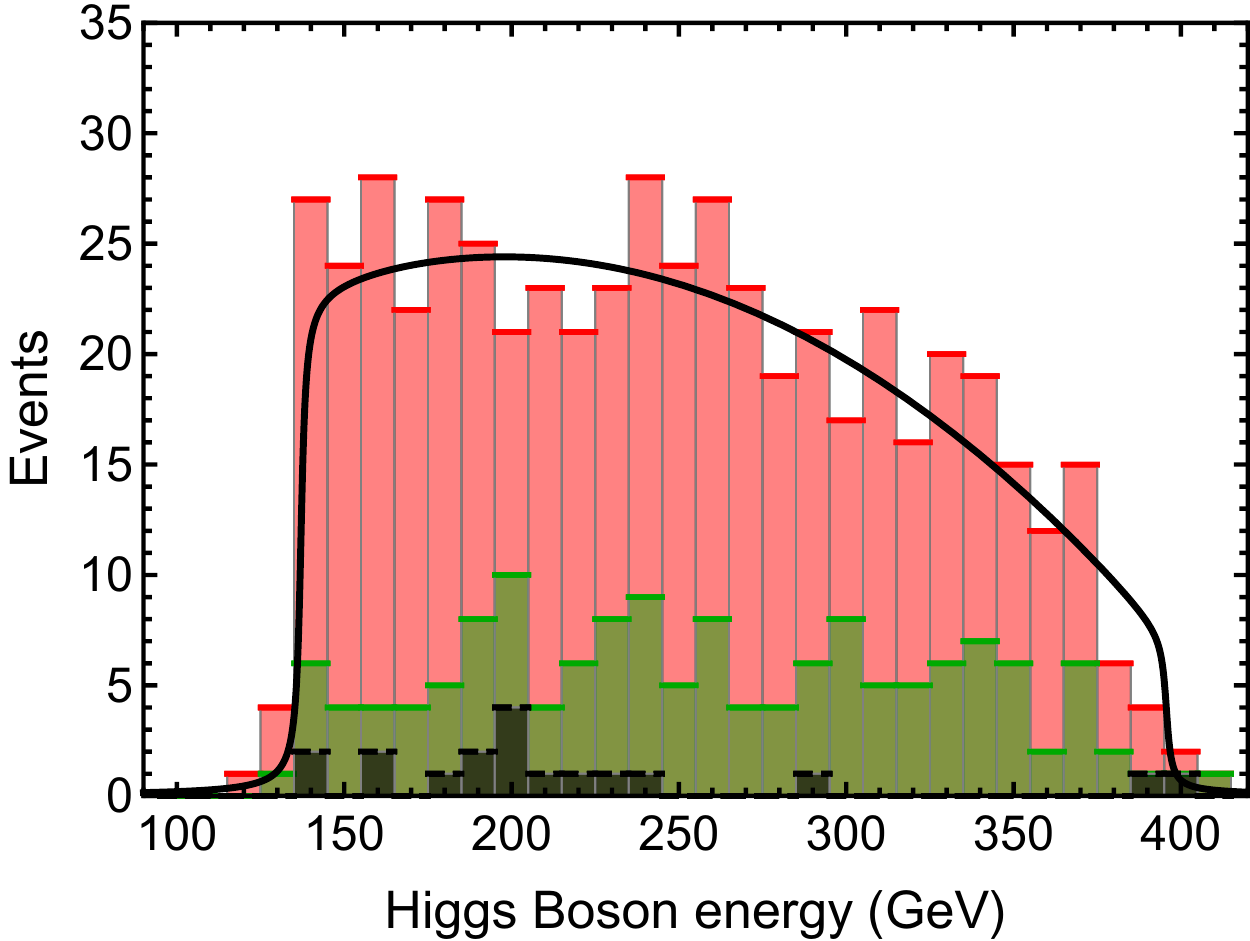}
\caption{Energy spectra of reconstructed $W$ (top left), $Z$ (top right), and $h$ bosons (bottom right) for Scenario DEG. SM background is shown in black, while $\tilde\nu_L$, and $\tilde\tau_1$ contributions are shown in green and blue, respectively. Cascade contributions from $\tilde e_{L,R}$, $\tilde\mu_{L,R}$, and $\tilde\tau_2$ are shown in red. The solid line is the result of the fit.}
\label{fig:energyC}
\end{figure}
Scenario DEG has degenerate soft slepton masses, meaning that in general we expect a much larger number of events, and thus better statistics. The boson energy spectrum and fit are shown in Fig.~\ref{fig:energyC}. Due to the vanishing flavor mixing, all $\tilde\nu_{L\ell}$ will decay in the same way, enhancing their contribution to $Z$-like and $h$-like events. Since only $\tilde\tau_1$ decays into on-shell $W^\pm$ bosons, the signal $W$-like events should be similar to the ST scenario.

In this scenario, both $\tilde e$ and $\tilde\mu$ will contribute to all datasets via their cascade decays. As in scenario SE, we expect that the endpoint analysis for $Z$-like and $h$-like events will still be valid in spite of these additional processes. However, the cascades also contribute to the $W$-like dataset from incorrectly identified $Z$ bosons, leading to the wrong reconstruction of the lower endpoint. In order to avoid this problem, we first use the standard endpoint method with the $Z$-like and $h$-like events in order to obtain $m_{\tilde\nu_L}$ and $m_{\tilde\nu_R}$. Then, as commented in Sec.~\ref{sec:method}, once we have the R-sneutrino mass, we can use only the upper endpoint to obtain $m_{\tilde\tau_1}$. This procedure gives us the three slepton masses with good agreement with the theoretical input.

\begin{table}[t]
\centering
\setlength{\tabcolsep}{1.5em}
{\begin{tabular}{| c | c | c|  c||c|} 
\hline
Scenario & SE & ST & DEG & Theory \\
\hline
$m_{\tilde\ell_{1}}$(GeV) & - & 296.91 $\pm$ 10.69& 290.51 $\pm$ 10.01&294.47\\
\hline
$m_{\tilde{\nu}_L}$ (GeV) & 293.63 $\pm$ 3.12 & 293.32 $\pm$ 3.61& 293.41 $\pm$ 2.15&293.37\\
\hline
$m_{\tilde{\nu}_R}$ (GeV) & 100.52 $\pm$ 1.65 & 101.14 $\pm$ 1.36& 100.05 $\pm$ 0.67&100.00\\
\hline
\end{tabular}}
\caption{Reconstructed masses in our three scenarios. For $m_{\tilde\ell_1}$, the last column shows the prediction for the lightest stau mass. }
\label{table:results}
\end{table}
We can then conclude that, even though originally designed for a simplified scenario with charginos and neutralinos, the endpoint method can be used to get a good first estimate of the lightest slepton masses of our model. This analysis has been carried out with only two $500$~fb$^{-1}$ samples of data, with type \textbf{B} and \textbf{L} polarizations respectively, which is small compared to the total 8~ab$^{-1}$ of integrated luminosity in the original proposal. 

\section{Conclusions}
\label{sec:conclusions}

We have investigated scenarios where sleptons decay into R-sneutrinos and a SM boson. Such scenarios are challenging for the LHC as the SM bosons decay dominantly into hadrons. We have found that current LHC data still allow for relatively light sleptons with masses below 200 GeV. In case that only one generation of sleptons is light, even the upcoming LHC with a luminosity of 300 fb$^{-1}$ cannot exclude such light sleptons even though the accessible parameter space gets constrained further. In case that all three generations of sleptons have about the same mass, the upcoming LHC run will push the mass limit to about 225~GeV.

We have then addressed the question to which extent a future ILC running at 1 TeV can discover such scenarios. Here we have found that sleptons with masses of up to 400 GeV can be discovered with a luminosity of 100 fb$^{-1}$ provided that $m_{\tilde l} - m_B - m_{\tilde \nu_R} \gsim 60$~GeV. In case that the luminiosity is increased to 1 ab$^{-1}$, sleptons with masses of up to 450 GeV can be discovered even if the allowed phases space is smaller. An important ingredient is the polarization of both $e^-$ and $e^+$ for a sufficient suppression of the SM background.

Last but not least, we have investigated how well the masses of sleptons
can be measured in such scenarios, assuming that their masses are in the ballpark of 300 GeV and the $\tilde \nu_R$ has a mass around 100 GeV. For this we have adapted an endpoint method developed for the mass measurements of neutralinos and charginos, in case that these decay dominantly into SM-bosons yielding similiar final states. As long as a slepton does not lead to a cascade decay, we find that the method can reconstruct its mass with a precision of a few percent.

\section*{Acknowledgements}

The authors would like to thank Jenny List for discussing the mass measurement method. W.P.\ has been supported by DAAD, project no.\ 57395885. N.C.V.\ was funded by grant No.\ 236-2015-FONDECYT. J.J.P.\ and J.M.~acknowledge funding by the {\it Direcci\'on de Gesti\'on de la Investigaci\'on} at PUCP, through grant No.\ DGI-2019-3-0044. N.C.V., J.J.P., and J.M. have been also supported by the DAAD-CONCYTEC project No.\ 131-2017-FONDECYT. 

\appendix

\section{Numerical Tools}

Throughout this paper we have used \texttt{SARAH 4.14.0}~\cite{Staub:2008uz,Staub:2013tta,Staub:2012pb,Staub:2010jh,Staub:2009bi} to implement the model in \texttt{SPheno 4.0.4}~\cite{Porod:2003um,Porod:2011nf}, which calculates the mass spectrum and branching ratios. We used \texttt{SSP 1.2.5}~\cite{Staub:2011dp} to carry out the parameter variation. The \texttt{SARAH} output also includes UFO files~\cite{Degrande:2011ua} that enter LHC and ILC event generators.

For LHC studies we use \texttt{MadGraph5\_aMC@NLO 2.7.0}~\cite{Alwall:2014hca} followed by \texttt{PYTHIA 8.244}~\cite{Sjostrand:2006za}, which generates the showering and hadronization. Events are generated with the \texttt{CTEQ6L1} PDF set~\cite{Pumplin:2002vw}. The detector simulation and event reconstruction is carried out by \texttt{DELPHES 3.4.2}~\cite{deFavereau:2013fsa,Cacciari:2011ma}, using the built-in ATLAS and CMS cards. To generate the exclusion regions we processed these events by by \texttt{CheckMATE 2.0.26}~\cite{Drees:2013wra,Dercks:2016npn}, which determines if a specific point has been excluded or not by the considered searches. 

For our ILC analysis we use \texttt{WHIZARD~2.6.2}~\cite{Kilian:2007gr,Moretti:2001zz}. This simulation includes ISR and beamstrahlung implemented with \texttt{CIRCE1-2.2.0}~\cite{Ohl:1996fi}. The parton shower and hadronization of the jets was carried out with the built-in version of \texttt{PYTHIA~6.427}. The detector simulation was again done by \texttt{DELPHES}, using the built-in ILD card.

\bibliographystyle{utphys}
\bibliography{SleptonSnuILC}

\end{document}